\documentclass{article}

\usepackage[a4paper, total={6.5in, 10in}]{geometry}
\usepackage[utf8]{inputenc} 
\usepackage[T1]{fontenc}
\usepackage{hyperref}       
\hypersetup{
    colorlinks=true,
    citecolor=blue,
    linkcolor=blue,
    filecolor=magenta,      
    urlcolor=cyan,
    pdftitle={Overleaf Example},
    pdfpagemode=FullScreen,
    }
\usepackage{url}            
\usepackage{booktabs}       
\usepackage{amsfonts}       
\usepackage{nicefrac}       
\usepackage{microtype}      
\usepackage{lipsum}
\usepackage{fancyhdr}       
\usepackage{graphicx}       
\graphicspath{{media/}}     
\usepackage{amsmath}        
\DeclareUnicodeCharacter{2212}{-}

\usepackage[labelfont=bf]{caption}

\usepackage{graphicx}
\usepackage{adjustbox}
\usepackage{subcaption}

\usepackage{mathtools}

\usepackage{bm} 
\usepackage{bbm} 
\usepackage{dsfont} 

\usepackage{algorithm}
\usepackage[noend]{algpseudocode}

\usepackage{xcolor}

\usepackage{tabularx}
\usepackage{multirow}

\usepackage{nicematrix}
\usepackage{calc}
\newcolumntype{P}[1]{>{\raggedright\arraybackslash}p{#1\textwidth-2\tabcolsep-1.5\arrayrulewidth}}
\newcolumntype{Y}[1]{>{\centering\arraybackslash}p{#1\textwidth-2\tabcolsep-1.5\arrayrulewidth}}

\usepackage{anyfontsize}
\usepackage{tikz}

\usepackage{colortbl}

\usepackage{array}
\usepackage{makecell}

\newcolumntype{M}[1]{>{\centering\arraybackslash}m{#1}}
\newcolumntype{N}{@{}m{0pt}@{}}

\usepackage{stackengine}

\usepackage{authblk}

\title{Modelling and Predicting Online Vaccination Views using Bow-tie Decomposition}
\author[1,2]{Yueting Han}
\author[2,3]{Marya Bazzi}
\author[4]{Paolo Turrini}
\affil[1]{\small MathSys CDT, University of Warwick, Coventry, UK}
\affil[2]{\small Mathematics Institute, University of Warwick, Coventry, UK}
\affil[3]{\small The Alan Turing Institute, London, UK}
\affil[4]{\small Department of Computer Science, University of Warwick, Coventry, UK}

\begin{document}

\date{}
\maketitle

\begin{abstract}
Social media has become increasingly important in shaping public vaccination views, especially since the COVID-19 outbreak. 
This paper uses bow-tie structure to analyse a temporal dataset of directed online social networks that represent the information exchange among anti-vaccination, pro-vaccination, and neutral Facebook pages.
Bow-tie structure decomposes a network into seven components, with two components ``SCC'' and ``OUT'' emphasised in this paper: SCC is the largest strongly connected component, acting as an ``information magnifier'', and OUT contains all nodes with a directed path from a node in SCC, acting as an ``information creator''.
We consistently observe statistically significant bow-tie structures with different dominant components for each vaccination group over time. In particular, the anti-vaccination group has a large OUT, and the pro-vaccination group has a large SCC. 
We further investigate changes in opinions over time, as measured by follower count variations, using agent-based simulations and machine learning models. 
Across both methods, accounting for bow-tie decomposition better reflects information flow differences among vaccination groups and improves our opinion dynamics prediction results.  
The modelling frameworks we consider can be applied to any multi-stance temporal network and could form a basis for exploring opinion dynamics using bow-tie structure in a wide range of applications. 
\end{abstract}

\section{Introduction}
\label{sec:intro}
Vaccination campaigns have drawn long-standing public attention \cite{Tafuri-2014, MacDonald-2015}, particularly since the outbreak of the COVID-19 pandemic \cite{Fernando-2020, Merryn-2020, Menni-2021, Sahil-2021}. Given the significant impact of online social media platforms as sources of information, a number of studies have emphasised their effect on vaccination views in public opinion \cite{Davies-2002, Evrony-2017, Burki-2019, Roozenbeek-2020, Wilson-2020, Fidelia-2022}. 
Recent studies have highlighted the significance of the information ``creator-receiver'' dynamics in online vaccination campaigns: 
some researchers have found that vaccination opponents tend to produce a higher volume of information than vaccination supporters \cite{Fidelia-2022, Johnson-2020, Illari-2022, Germani-2021}; 
several studies have observed that most (mis)information is created by a minority of users (which should not be assumed to be representative of a majority) and that information roles tend to remain relatively stable over time \cite{Castioni-2022, Germani-2021}. 

Building on these previous studies, this paper explores online behavioural differences among vaccination groups, namely vaccination supporters, opponents, and neutrals.
Instead of simply dividing online users into two categories (i.e., ``creators'' and ``receivers'') based on the volume of messages they create or receive, this paper explores a more nuanced division of roles each user might play in online information flow using a network structure called ``bow-tie structure''. It was recently explored in the context of online debates \cite{Mattei-2022}, and we introduce it next. 

\begin{figure}
    \centering
    \adjincludegraphics[width=\textwidth,trim={{.013\width} {.1\height} {.01\width} {.1\height}},clip]{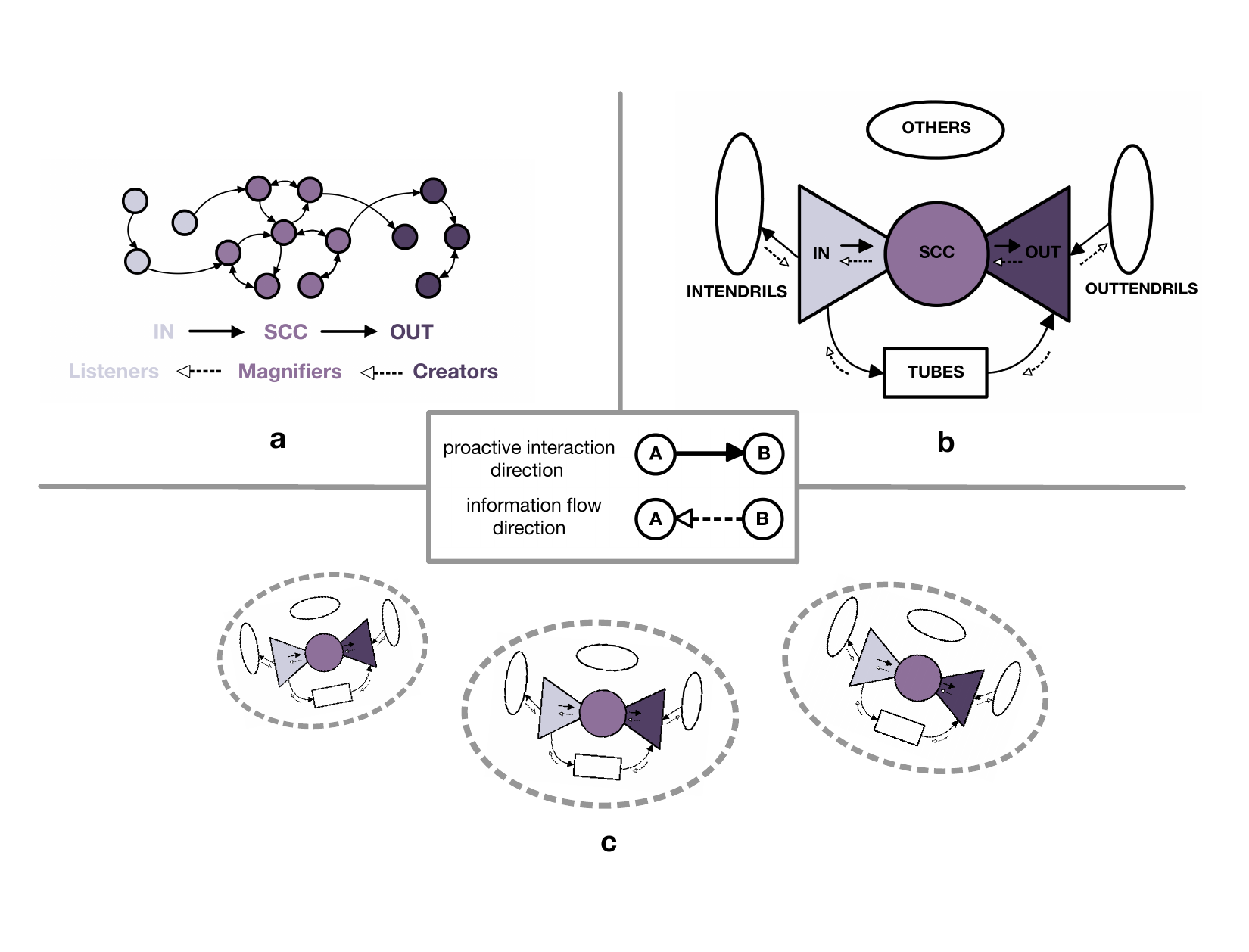}
    \caption{\textbf{Bow-tie structures in online social networks.} The arrows highlight that in this paper, an edge from node A to B in online social networks (solid arrow) represents an interaction from page A to B (e.g., page A recommends page B to its members), while the direction of information flow (dashed arrow) goes in the opposite direction (e.g., content about page B is presented or ``flows'' to page A).
    \textbf{a, primitive bow-tie structure.} This panel illustrates a bow-tie structure that divides a toy example network into three components: SCC, IN, and OUT. This decomposition establishes pairwise relations between these components, assigning distinct roles to each in terms of information flow: IN - ``listeners'', SCC - ``magnifiers'', and OUT - ``creators''. 
    \textbf{b, extended bow-tie structure.} This panel expands on the bow-tie structure in panel a by introducing additional components: TUBES, INTENDRILS, OUTTENDRILS, and OTHERS. In this structure, IN not only ``listens'' to SCC but also to INTENDRILS and TUBES, while OUT not only ``creates'' information that is delivered to SCC but also to TUBES and OUTTENDRILS.
    \textbf{c, recursive bow-tie structure.} This panel displays an example of a recursive bow-tie structure, where the entire graph is partitioned into subgraphs, and bow-tie decomposition is applied to each of them. Note that edges across partitioned subgraphs are disregarded in this case.}
    \label{fig:intro}
\end{figure}

\paragraph{Bow-tie structure.} Bow-tie structure was introduced by Broder et al. \cite{Broder-2000} in 2000 as a type of network structure that encodes the connectivity of the World Wide Web (WWW), with nodes representing pages and edges representing hyperlinks. 
The primitive form of bow-tie structure divides a directed network into four components: the largest strongly connected component (SCC), the in-periphery component (IN) which includes all nodes with a directed path to a node in SCC, the out-periphery component (OUT) which comprises all nodes with a directed path from a node in SCC, and the other sets (OTHERS) for all remaining nodes. The first three components of bow-tie structure (see Figure \ref{fig:intro}a) are shaped like a bow-tie, with SCC acting as the knot and IN and OUT components as the fans. 
Bow-tie structure was later refined by Yang et al. \cite{Yang-2011} through the introduction of TUBES, INTENDRILS, and OUTTENDRILS, as shown in Figure \ref{fig:intro}b. Yang et al. also proved that any directed graph can be decomposed into a bow-tie structure.

Online social media can be viewed through a similar lens to that of the World Wide Web (WWW), where pages are run by social media users and directed interactions exist between pages, such as thumbs-up, reposts, and following \cite{Obar-2015}. 
In this paper, we define an edge from node A to B in an online social network as an interaction from page A to page B (e.g., page A recommends page B). Such an interaction is often triggered by page B presenting some content of interest to page A. 

In the light of this, bow-tie structure can provide useful insight into interpreting the various ``role'' of pages in online information flow \cite{Mattei-2022, Zhang-2007, Elliott-2020}. Pages in SCC are usually active in two-way interactions (e.g., sharing strong communicative interests), and thus may be regarded as information ``magnifiers''. OUT pages are often passively interacted with by SCC, INTENDRILS, and TUBES pages but seldom engage in directed interactions with these pages. This suggests a high tendency of OUT pages to present their content to the public, behaving as information ``creators''. IN pages generally act in the opposite way, serving as information ``listeners'' to pages in SCC, OUTTENDRILS, and TUBES. OTHERS usually contains pages with sparse interactions with most pages in other bow-tie components, either proactively or passively. For finer-scale analysis of such information flow, recursive bow-tie structures have been explored by considering bow-tie structures of subgraphs, in contrast to the entire network \cite{Mattei-2022, Fujita-2019} (see an illustration in Figure \ref{fig:intro}c).
Subgraphs are often extracted through the application of computational methods \cite{Mattei-2022, Fujita-2019}, such as community detection where ``densely connected'' users are grouped together \cite{Fujita-2019}. Manual examination may also be employed to ensure users sharing specific discussion topics are appropriately grouped \cite{Mattei-2022}.  

To investigate bow-tie structure in online vaccination campaigns, we study a real-world temporal dataset about an online vaccination campaign, which is publicly available and was previously analysed in 2020 by Johnson et al. \cite{Johnson-2020}. It describes two snapshots of online recommendations between Facebook pages in February and October 2019 (before the COVID-19 outbreak), with each page manually checked and assigned a vaccination stance of ``anti'', ``pro'', or ``neutral''\footnote{Further details on stance annotation can be found in \cite{Johnson-2020, Illari-2022}. Here, neutral pages ``focus around vaccines or another topic (e.g., a school parent association that has become linked to the vaccine debate but for which the stance is still undecided)'', as introduced by Johnson et al. \cite{Johnson-2020}.}. 
The dataset was originally analysed based on page-level interactions, the number of members who subscribe to each page (i.e., the ``fan size''), narratives (e.g., safety concerns and conspiracy theories), and geography. Information on geography and narratives was not made public by the authors and is currently not available. 
Data on vaccination stance, time stamps, and fan size allows us to investigate the following important questions about bow-tie structures: 
(1) explanatory power, i.e., whether the stance on vaccination is associated with a different bow-tie structure, and how to explain any existing differences; 
(2) temporal stability, i.e., whether bow-tie structures remain stable through time; 
(3) predictability, i.e., whether bow-tie structures can help predict fan size variations, serving as a reflection of the dynamic nature of online vaccination views\footnote{Follow-up research \cite{Illari-2022} provides an extended version of this dataset with an additional two snapshots -- November 2019 and December 2020 (during the initial stage of the COVID-19 pandemic), although without time-stamped fan size data. This limits our bow-tie analysis. Therefore, we focus only on the original dataset in \cite{Johnson-2020} in our main paper and present some results of the extended dataset in Supplementary Material.}.

\paragraph{Contribution.}
This paper uses bow-tie decomposition to analyse and predict online vaccination views. We build on  Mattei et al.'s contribution \cite{Mattei-2022}, which identifies different recursive bow-tie structures in online social networks and apply this idea to the online vaccination views dataset from Johnson et al. \cite{Johnson-2020, Illari-2022}. 
To our knowledge, there is very limited prior research applying bow-tie structure to investigate and predict the spread of opinion dynamics in social media networks. 
Additionally, while Johnson et al. \cite{Johnson-2020} provide an opinion dynamics prediction model on the same dataset that relies solely on page fan counts and disregards network structures, our study 
incorporate bow-tie structure into the modeling framework.

The contribution of this paper is twofold: firstly, we find that online vaccination groups (i.e., pages holding anti-vaccination, pro-vaccination, and neutral viewpoints) exhibit different bow-tie structures, which we interpret in the light of information flow roles. Secondly, using agent-based epidemic simulations and machine learning models, we explore how these structures reflect information flow differences among vaccination groups and their potential to predict online vaccination view dynamics as quantified by fan size page variation. 
The modelling frameworks we use are general and could form a basis for exploring opinion dynamics using bow-tie structure in a wide range of applications.

\paragraph{Paper Structure.} 
This paper is organised as follows.
In \hyperref[sec:dataset]{Section 2}, we describe the online recommendation dataset about vaccination views.
\hyperref[sec:method]{Section 3} outlines the methodology of bow-tie decomposition in this paper.
In \hyperref[sec:results]{Section 4}, we present our findings by first detecting and interpreting bow-tie structures in this dataset, and secondly using this structure in agent-based and machine learning models to predict dynamics in online vaccination views. 
Finally, \hyperref[sec:discussion]{Section 5} summarises our main results and discusses directions of future work.

\section{Data Description}
\label{sec:dataset}
\begin{figure}
    \centering
    \adjincludegraphics[width=\textwidth,trim={{.04\width} {.22\height} {.02\width} {.19\height}},clip]{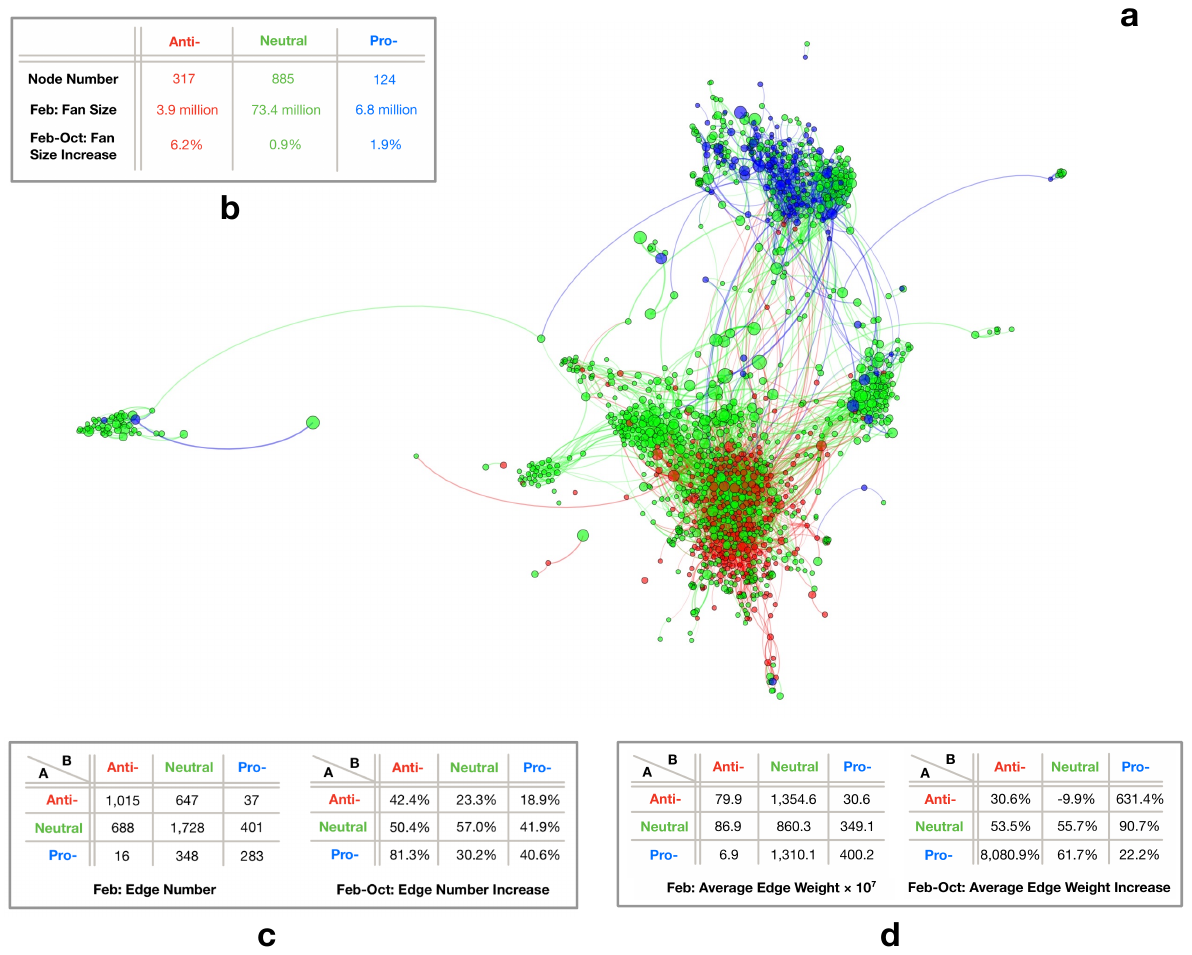}
    \caption{\textbf{Online recommendation networks about vaccination views. a, February network.} This is a snapshot of the largest weakly connected subgraph in February 2019, reproduced from Johnson et al.'s paper \cite{Johnson-2020}. It includes over 94\% of nodes and 99\% of edges from the entire network. Each node represents a page. Its node colour depicts page polarity: red for anti,  green for neutral, and blue for pro. Its node size is proportional to its page fan size. The node layout follows ForceAtlas2 in Gephi.
    The edge colour follows the colour of its source node (where the edge starts from).
    \textbf{b, node-level data. }It describes the total number of nodes and the total fan size for each vaccination group. By observation, the neutral vaccination group dominates with the largest number of pages and fans. The pro-vaccination group has fewer pages but a stronger fan base than the anti-vaccination group, mainly due to three pages with over a million fans. The anti-vaccination group has no pages with over a million fans in February and October but experienced the largest percentage increase in fans from February to October.
    \textbf{c, d, edge-level data. }It describes the edge number (panel c) and the average edge weight (panel d) within and across vaccination groups. Every edge is directed from A to B (i.e., A recommends B). It can be observed that the direction and weight of recommendations are important. Despite that, there are a larger number of edges within vaccination groups than across vaccination groups (except for pro- pages), and the highest edge weights flow from anti- to neutral and pro- to neutral groups (possibly due to the neutral group's high activity in interacting with both groups). Additionally, the anti- and pro-vaccination groups had minor interaction in February but interestingly experienced drastic increases in both edge number and average weight from February to October.
    }
    \label{fig:data}
\end{figure}

The dataset from Johnson et al. \cite{Johnson-2020} consists of two snapshots of online competition between different vaccination views in February and October 2019, involving nearly $100$ million users on Facebook across countries, continents, and languages. It can be represented by two directed networks corresponding to February and October. 
The February network is illustrated in Figure \ref{fig:data}a.
The number of nodes is the same in February and October, given by $1326$ in total (see further data in Figure \ref{fig:data}b). The number of edges in February is $5163$, and in October is $7484$ (see further data in Figure \ref{fig:data}c). The original data was in .pdf form, we pre-process it and make it easily accessible in a variety of analysis-ready formats on \href{https://github.com/YuetingH/BT_Vaccination_Views}{GitHub}.

Details are explained below:
\begin{itemize}
    \item \textbf{Node: }Each node represents a public Facebook page that discusses vaccination topics. It is attributed with fan size, that is, the number of members who subscribe to the Facebook page, along with the other attribute polarity including anti-vaccination, pro-vaccination and neutral. 
    Here, neutral pages ``focus around vaccines or another topic (e.g., a school parent association that has become linked to the vaccine debate but for which the stance is still undecided)'', as introduced by Johnson et al. \cite{Johnson-2020}.
    Whereas its polarity remains the same for February and October snapshots, its fan size can either increase or decrease. 
    
    Remarks: (1) For consistency, a red, blue, or green node will always represent the page in the anti-, pro-vaccination, and neutral group when it comes to analysing this dataset. (2) It is allowed for a page to have no fans (this is the case for a total of 4 pages in February and 10 pages in October).
    
    \item \textbf{Edge: }A directed edge from node A to B means page A recommends B to all its members at the page level, as opposed to a page member simply mentioning another page. The number of times a page is recommended and the exact timestamp of the recommendation is not recorded in the dataset. Rather, an edge from A to B is present in a given monthly snapshot if page B was recommended to all members of A at some point earlier or within the month. In that sense, the network represents the cumulative recommendations over time, with edges in February also appearing in October, but not the other way around. 
    Note that, despite the cumulative nature of the edges, the differences between the two snapshots are evident and visually shown in \cite{Johnson-2020} Figure 2a.
    
    Remarks: Both ``two-way recommendation'' and ``self-recommendation'' are allowed. Two-way recommendation means two pages recommend each other (involving $8.5\%$ of all recommendations in February and $10.2\%$ in October).
    Self-recommendation, where a page recommends itself to all of its fans (e.g., to increase engagement), is rare in our dataset (less than 0.2\% in both February and October).
\end{itemize}

During preprocessing, we define an \textbf{edge weight} to quantify the significance of each recommendation (see Figure \ref{fig:data}d). It is obtained by the product of both ends' fan size. 
This edge weight choice builds on the intuition that more fans may be recruited if pages on both ends have larger fan sizes. On the one hand, being recommended to another page with a larger fan size will likely attract more fans. On the other hand, a recommended page with a larger fan size is potentially more influential to other pages; thus, more fans may be recruited accordingly.
This weight choice is also mentioned as a ``product kernel'' by Johnson et al. in their prediction model \cite{Johnson-2020}, which was previously shown to be useful in Palla et al. \cite{product-kernel}.
Although bow-tie structure disregards edge weight, community detection involved in recursive bow-tie structure can incorporate this factor, so as to produce a better partition of the network. See details in \hyperref[sec:methodrbt]{Section 3(b)}.

\section{Methodology}
\label{sec:method}
We focus on a particular network type throughout this paper, aligned with our dataset: directed weighted networks with self-loops and without multi-edges.
Some preliminary graph definitions are listed below. 

\begin{itemize}
    \item A \textit{network} $G = (V, A)$ consists of a set of nodes $V$ and a weighted adjacency matrix $A = (A_{u,v})_{u,v \in V}$. $A_{u,v} = w > 0$ if there is an edge from node $u$ to $v$ with weight $w$ and $A_{u,v} = 0$ otherwise.
    
    \item A \textit{path} from $u \in V$ to $v \in V$ is defined as a succession of nodes $(n_0, n_1, ..., n_k)$, where $k$ is a nonnegative integer, $n_0 = u$, $n_k = v$, and for any $i = 1, ..., k$, $n_i \in V$ are distinct satisfying $A_{n_{i-1},n_{i}} > 0$. As a special case, every node $u \in V$ is considered to have a path $(n_0 = u)$ to itself.
    
    \item Let $u, v \in V$ be nodes and $T \subseteq V$ be a subset of nodes. A node $v$ is said to be \textit{reachable} from $u$ if a path exists from $u$ to $v$. As an extension, $u$ is said to be reachable from $T$ if there exists at least a $w \in T$ such that $u$ is reachable from $w$. $T$ is said to be reachable from $u$ if there exists at least a $w \in T$ such that $w$ is reachable from $u$. 


    \item A \textit{subgraph} $G' = (V', A')$ of a graph $G = (V, A)$ is a graph such that $V' \subseteq V$ and $A'$ satifies $A'_{u,v} = A_{u, v}$ for any $u, v \in V'$.

    \item  A \textit{strongly connected component} of a graph $G$ is a subgraph of $G$ where there exists a path from every node to every other node. 

    \item A \textit{hard partition} of a graph $G$ is a division of the set of nodes $V$ into $k$ non-overlapping sets $C_i$, $i = 1, 2, ..., k$, such that $\bigcup_{i=1}^{k} C_{i} = V$ and $C_i \cap C_j = \emptyset$ for any $i \neq j$. One can also consider soft partitions, where sets can overlap, but this is beyond the scope of this paper. We refer to a ``hard partition'' as a ``\textit{partition}'' throughout the paper. 
\end{itemize}

\subsection{Bow-tie Structure}
In this paper, we adopt the bow-tie structure definition and detection algorithm presented by Yang et al. \cite{Yang-2011}, also used in Mattei et al. \cite{Mattei-2022}.

\paragraph{Definition.}
Assume $S$ is the largest strongly connected component of $G$. 
Bow-tie structure of $G$ consists of the following sets of nodes:
$$ \text{SCC} = S,\quad \text{IN} = \{v \in (V - S)\,|\,S\text{ is reachable from }v\}\footnotemark,\quad \text{OUT} = \{v \in (V - S)\,|\,v\text{ is reachable from }S\}$$
$$ \text{TUBES} = \{v \in (V - S - \text{IN} - \text{OUT})\,|\,v\text{ is reachable from IN and OUT is reachable from }v\} $$
$$ \text{INTENDRILS} = \{v \in (V - S)\,|\,v\text{ is reachable from IN and OUT is not reachable from }v\} $$
$$ \text{OUTTENDRILS} = \{v \in (V - S)\,|\,v\text{ is not reachable from IN and OUT is reachable from }v\} $$
$$ \text{OTHERS} = V - S - \text{IN} - \text{OUT} - \text{TUBES} - \text{INTENDRILS} - \text{OUTTENDRILS} $$
\footnotetext{For any sets $A$ and $B$, we define $A − B$ as a set such that for any $i \in A - B$, $i \in A$ and $i \notin B$.}

Yang et al. \cite{Yang-2011} also proved that the sets of bow-tie components above are mutually disjoint and thus form a partition of the nodes. In other words, any directed graph can be decomposed into a bow-tie structure.  

\paragraph{Algorithm.}
The detection of the largest strongly connected component is a well-established process in the field of graph theory, with early works from \cite{Tarjan-1972, Nuutila-1994}.
The algorithm for obtaining the remaining bow-tie components is outlined by Yang et al. \cite{Yang-2011}.  
The entire algorithm for detecting bow-tie structure has been implemented in code on \href{https://github.com/alan-turing-institute/directedCorePeripheryPaper/blob/master/bowtie_detect.py}{GitHub}, and we employ the same code for our analysis.

\subsection{Recursive Bow-tie Structure}
\label{sec:methodrbt}

\paragraph{Background.} Bow-tie structure of an entire network is largely dependent on the generation of its edges, which, however, often exhibit some amount of randomness in online social networks \cite{Illari-2022, Ying-2009}.
For instance, users may randomly interact with recommended strangers \cite{Ying-2009, Lingam-2020}. Social bots, creating fake accounts and spreading spam, also contribute to this randomness \cite{Lingam-2020, Chu-2012}. 
This has motivated researchers to move beyond the potential randomness of edges and delve deeper into the analysis of bow-tie structure on a finer scale, also referred to as ``recursive bow-tie structure''. 

Recursive bow-tie structure has been explored by considering partitioned subgraphs or different choices of SCC. 
The first considers bow-tie decomposition on partitioned subgraphs while ignoring inter-subgraph edges. 
This approach is grounded on the empirical observation that online social networks usually exhibit community structures, where nodes within a set are densely connected and inter-set connections are relatively sparse \cite{Fujita-2019, Lingam-2020, Dill-2002}. 
These communities are often composed of users who share common communicative goals, referred to as ``discursive communities'' by Mattei et al. \cite{Mattei-2022}. For instance, in their research, bow-tie decomposition is applied to communities including left/right politicians and official accounts of governments and media (e.g. newspapers, TV channels, and journalists) examined through metadata. Another example from Fujita et al. \cite{Fujita-2019} employs a computational technique to perform bow-tie analysis on densely connected communities. Bow-tie role assignments produced by this type of recursive bow-tie structure are interpretable, as they reflect the local roles of nodes within each subgraph.
The second type of recursive bow-tie structure considers other strongly connected components rather than the largest one \cite{Yang-2011, Lu-2016, paolo}. This second method, while theoretically feasible, is often less interpretable as it may assign multiple roles to the same nodes.

\paragraph{Algorithm.} This paper focuses solely on the first approach (see its implementation in \hyperref[algorithm:rbt]{Algorithm 1}), which considers bow-tie structures of partitioned discursive communities as described by Mattei et al. \cite{Mattei-2022}. Associated with our online vaccination view dataset, we interpret the discursive communities in two ways: (1) vaccination groups (anti-, pro-, neutral), and (2) densely connected communities detected by computational techniques, where the number or size of communities is not fixed.
As a result, two kinds of decomposition will be implemented for each network snapshot, enabling intra (i.e., approach (1)) and inter (i.e., approach (2)) vaccination groups' bow-tie analysis over time.

Note that while subgraphs of vaccination groups can be easily extracted based on the metadata from Johnson et al. \cite{Johnson-2020}, community detection typically requires the use of computational heuristics, with multiple choices available \cite{comm_newman, consensus, comm, comm-literaturereview, comm-literaturereview2}.  
For the purposes of this paper, we use a widely used community detection method known as Infomap. This approach leverages Shannon Entropy and random walks with edge weights for community detection \cite{comm}. 
It introduces some stochasticity to the identified partition, but our analysis indicates that it has a minor influence on our results. 
Infomap may also detect small communities with fewer than five nodes, but these are less meaningful in our context. To address this, we label nodes in such communities as ``UNASSIGNED''. 
We provide further details in Supplementary Material to explain why we choose Infomap over modularity maximisation \cite{comm_newman} (another popular community detection method) for our specific problem, and examine the stochasticity of Infomap in our dataset.

\begin{algorithm}[bpht]
\label{algorithm:rbt}
\caption{Recursive Bow-tie Detection}
\hspace*{\algorithmicindent} \textbf{Input:} graph $G$, partition $C = \{C_1, ..., C_k\}$ \\
\hspace*{\algorithmicindent} \textbf{Output:} $\text{SCC}^{\text{r}}$, $\text{IN}^{\text{r}}$, $\text{OUT}^{\text{r}}$, $\text{TUBES}^{\text{r}}$, $\text{INTENDRILS}^{\text{r}}$, $\text{OUTTENDRILS}^{\text{r}}$, $\text{OTHERS}^{\text{r}}$
\begin{algorithmic}[1]
\State For $i = 1, ... k $, construct a subgraph $G_i = (C_i, A_i)$ of the graph $G$.
\State For each subgraph $G_i$, acquire its bow-tie structure $\text{SCC}_{i}$, $\text{IN}_{i}$, $\text{OUT}_{i}$, $\text{TUBES}_{i}$, $\text{INTENDRILS}_{i}$, $\text{OUTTENDRILS}_{i}$, $\text{OTHERS}_{i}$. 
\State Obtain $\text{SCC}^{\text{r}} = \bigcup_{i=1}^{k} \text{SCC}_{i}$. Repeat this step to obtain other bow-tie components $\text{IN}^{\text{r}}$, $\text{OUT}^{\text{r}}$, $\text{TUBES}^{\text{r}}$, $\text{INTENDRILS}^{\text{r}}$, $\text{OUTTENDRILS}^{\text{r}}$, $\text{OTHERS}^{\text{r}}$.
\end{algorithmic}
\end{algorithm}

\section{Results}
\label{sec:results}
Our study focuses on two aspects: the detection of bow-tie structures in recommendation-based online social networks, and the analysis of these structures 
to enhance the prediction of opinion dynamics (modelled as page fan size over time).
Section 4.1 and Section 4.2 will discuss these two aspects, respectively.
\hyperref[sec:resultsBT]{Section 4(a)} explores whether the stance on vaccination is associated with different bow-tie structures, examines the stability of such structures over time, and assesses how one can interpret these observations.
\hyperref[sec:resultsOD]{Section 4(b)} uses and compares two approaches to predict page fan count variation: (1) supervised machine learning and (2) mechanistic simulation via agent-based epidemic models on information cascade.

As mentioned in \hyperref[sec:methodrbt]{Section 3(b)}, we consider bow-tie structures of ``discursive communities'' in the February-October 2019 network snapshots following Mattei et al. \cite{Mattei-2022}, where bow-tie structure is identified in subgraphs of the network (see Figure \ref{fig:intro}c). We characterise discursive communities in two ways that can yield complementary insights: (1) their view, in our case pro-, anti- and neutral-vaccination groups; (2) their placement in densely connected sets within the network, as defined by community structure which we detect using the commonly used flow-based method Infomap~\cite{comm}.
In this sense, two kinds of bow-tie decomposition are implemented for each network snapshot, enabling intra and inter vaccination groups bow-tie analysis. 
Each page is assigned a dual bow-tie role (different or not, such as SCC-SCC, SCC-OUT) at each timestamp. 
For clarity in our explanations, we refer to the first coarse-graining of discursive communities (i.e., when subgraphs are vaccination groups) as ``within-group bow-tie structure'', and a page's assigned role as its ``within-group bow-tie role''. On the other hand, we term the second coarse-graining of discursive communities (i.e., when subgraphs are communities identified with Infomap) as ``across-group bow-tie structure'', and a page's assigned role as its ``across-group bow-tie role''.


\subsection{Bow-tie Structure Detection in the Recommendation Networks}

\label{sec:resultsBT}

\begin{figure}
\centering
\adjincludegraphics[width=\textwidth,trim={{.11\width} {.03\height} {.13\width} {.02\height}},clip]{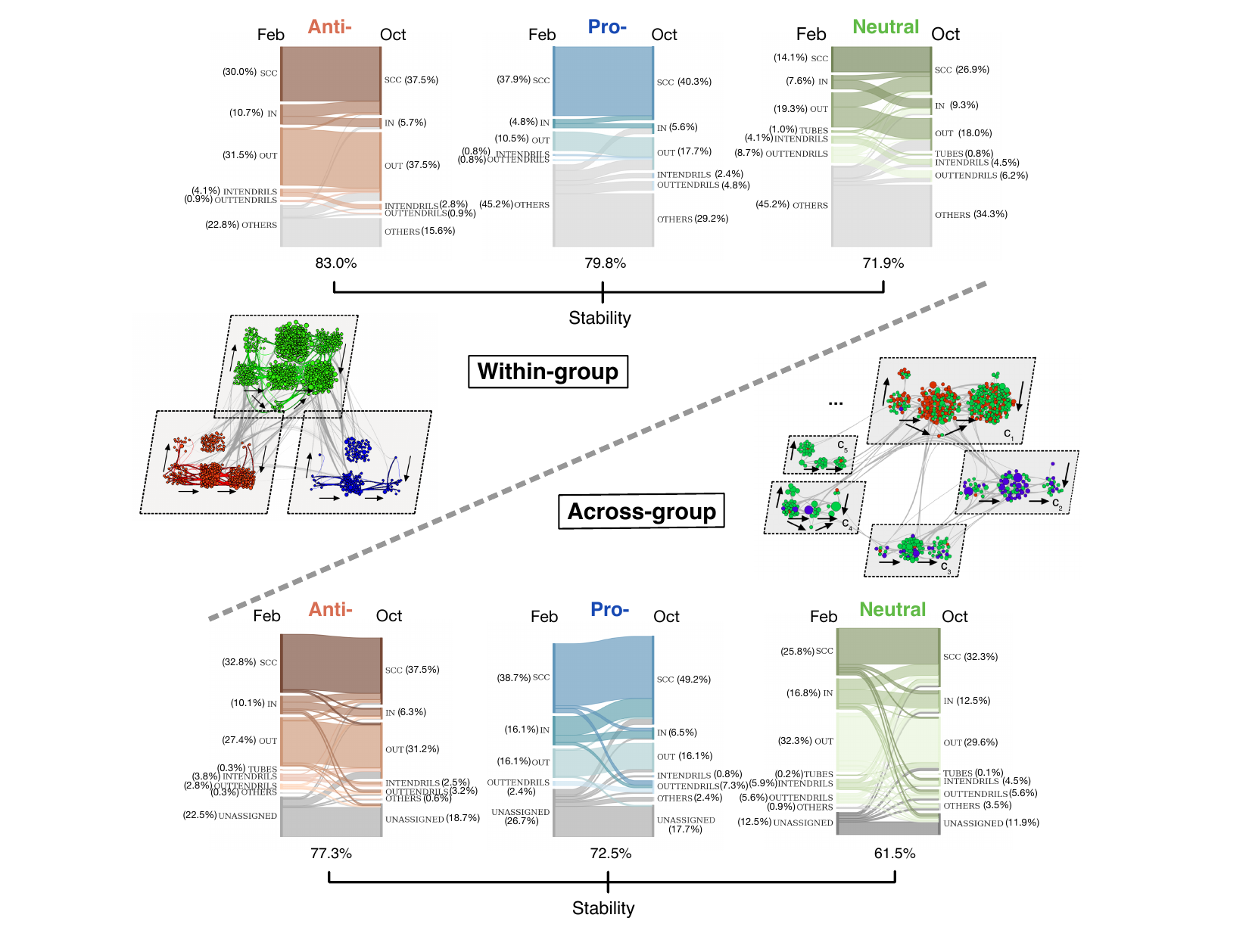}
\caption{\textbf{Within-group and across-group bow-tie structures in the February and October 2019 online recommendation-based networks.} 
The figure displays within-group bow-tie structures in the upper part and across-group bow-tie structures in the lower part for networks at both timestamps.
Each part includes \textbf{an explanatory diagram} of the decomposition scheme, where the February network is divided into subgraphs, and the bow-tie structure within each subgraph is revealed using an organised layout of node and arrow, maintaining consistency with Figure \ref{fig:intro}b. Nodes are colour-coded by vaccination group and proportionally sized based on fan size, consistent with Figure \ref{fig:data}a. 
Note that, the across-group diagram shows the largest five communities, labelled as $C_i$, ranked by node counts, with larger communities having smaller indices, which collectively represent $49.4\%$ of all pages. While all five largest communities are primarily composed of neutral pages, communities $C_1$ and $C_2$ stand out with nearly half of their pages anti- and pro-vaccination, respectively.
\textbf{Three Sankey diagrams} in each part illustrate the bow-tie structures for pages with different vaccination views. 
Each diagram has two columns representing bow-tie roles for February and October, with the flow indicating role variations\protect\footnotemark.
The stability, indicated beneath each diagram, quantifies the percentage of pages that maintain the same bow-tie roles at both timestamps.
Overall, these results indicate that the pro-vaccination group exhibits a large SCC in bow-tie decomposition for both choice of discursive communities, while the anti-vaccination group is comparatively dominated by OUT component. Moreover, these structures are stable over time. 
In contrast, the neutral group yields inconsistent bow-tie structures and exhibits less temporal stability.
}
\label{fig:obs}
\end{figure}
\footnotetext{In the across-group bow-tie structure, pages within communities containing fewer than 5 nodes were labelled ``UNASSIGNED'', as bow-tie structure of small communities may not be meaningful in our context. See details in \hyperref[sec:methodrbt]{Section 3(b)}.} 

Our results in Figure \ref{fig:obs} indicate that bow-tie structures associated with the pro-vaccination group consistently exhibit a large SCC in either way of bow-tie decomposition and at both February and October timestamps, whereas those associated with the anti-vaccination group have comparatively a large OUT component. 
In contrast, the neutral group demonstrates inconsistent results, with large OTHERS in within-group bow-tie decomposition, but a dissimilar pattern in across-group bow-tie decomposition with comparatively large main bow-tie components (i.e., SCC, OUT, and IN). 
Furthermore, anti-vaccination and pro-vaccination pages display a higher temporal stability in their bow-tie structures than neutral pages when transitioning from February to October, with anti-vaccination pages being slightly more stable than pro-vaccination pages. 

We offer some interpretations of the detected bow-tie structures. The large OUT component of the anti-vaccination group suggests a strong commitment to generating information. The pro-vaccination group's large SCC component underscores its strong information dissemination capability. Neutral pages have limited interactions with the ``mainstream'' (i.e., SCC, OUT, IN components) and are assigned the ``OTHERS'' role in the within-group bow-tie structure. In contrast, their interactions with anti- and pro-vaccination pages are more likely to focus on vaccination topics, resulting in strong across-group bow-tie structures with relatively large SCC, OUT, and IN components. Finally, in this dataset, pages in marginal components (i.e., OTHERS, INTENDRILS, OUTTENDRILS, and TUBES) tend to transition towards main components (i.e., SCC, OUT, and IN) over time. This shift may reflect increased integration into mainstream discussions, either through referencing SCC pages or being referenced by them.

Interestingly, while these observations are obtained using a different methodology (bow-tie structure), several are consistent with those discussed in Johnson et al. \cite{Johnson-2020} using node metadata. For example, the authors in \cite{Johnson-2020} mention that anti-vaccination groups generate a diverse range of narratives that blend topics such as ``safety concerns'' and ``conspiracy theories''. Furthermore, the pro-vaccination pages in the dataset are more centralised geographically than the anti-vaccination pages, which can facilitate reciprocal recommendations and amplify information dissemination. Neutral pages amongst themselves touch on a variety of topics (e.g., parenting and pets pages \cite{Johnson-2020, Illari-2022}) rather than mostly vaccination topics, which may help explain why they belong to ``OTHER'' when discursive communities are associated with vaccination stances. 

We end this section with observations on the robustness of our results. 
Firstly, we find that the detected bow-tie structures are statistically significant through comparisons with appropriately generated random graphs.
More details can be found in Supplementary Material.
Secondly, the follow-up dataset with two additional snapshots (i.e., November 2019 and December 2020) yields recursive bow-tie analysis results that are generally consistent with our findings in the dataset presented previously. Again, see more details in Supplementary Material. 
Thirdly, we observe that the temporal stability of across-group bow-tie decomposition is overall lower than that of within-group decomposition. This is not surprising, in the sense that the higher and varying number of detected communities compared to the constant small number of vaccination groups may favour more variation between snapshots. 


\subsection{Experiments with Detected Bow-tie Structure on Opinion Dynamics Prediction}
\label{sec:resultsOD}

Next, we examine whether one can use the February 2019 recommendation network to help predict fan count disparities for each individual page between February and October 2019, as a way to gain insight into opinion dynamics during this period. 
These predictions are implemented using supervised machine learning models and agent-based SIR models, with a specific emphasis on exploring whether bow-tie structure can improve these predictions. 

It is important to note that, a priori, one might expect recommendations to be more strongly correlated with fan size increase, than with fan size decrease. In other words, pages that receive recommendations are potentially likely to witness an increase in their fan counts over a short time frame, while those not recommended may not necessarily experience a decline. We come back to this point when discussing our results.   

Furthermore, we establish two assumptions below to centralise our research focus and enhance the interpretability of our results:
(1) We exclusively perform predictions on anti- and pro-vaccination pages, excluding neutral pages (while not ignoring the interactions of anti- and pro-pages with neutral pages). 
This choice is motivated by the fact that vaccination topics may not constitute the primary focus of neutral pages, such as those centred around parenting and pets \cite{Johnson-2020, Illari-2022}, and their fan count fluctuations may have a limited connection to recommendation networks concerning vaccination views; 
(2) Instead of covering all bow-tie components, we narrow our analysis to three key bow-tie components: SCC, OUT, and IN, while grouping the remaining components as ``NA''. 
This decision is made based on the following reasons.
Components SCC, OUT, and IN offer clearer insights for interpreting the roles of bow-tie components in information propagation. For instance, IN pages ``listen'' to SCC, INTENDRILS, and TUBES pages (main listeners), while OUTTENDRILS pages ``listen'' to OUT pages only (marginal listeners).
In addition, our empirical observations above show that the page count in SCC and OUT components effectively distinguishes between anti-vaccination and pro-vaccination pages, whereas other components lack such clarity.


\subsubsection{Supervised Machine Learning}

We begin by extracting categorical and numeric features of the anti- and pro-vaccination pages, including those related to bow-tie structure, from the February 2019 recommendation network snapshot, which potentially contribute to variations in page fan counts compared with October 2019 (see Table \ref{tab:feature}). 

\begin{table}[htbp]    
    \centering
    \begin{NiceTabular}{Y{0.26}P{0.31}Y{0.14}|P{0.31}}
    \toprule
    \fontsize{9pt}{9pt}\selectfont\textbf{Notation}  & \fontsize{9pt}{9pt}\selectfont\textbf{Description} & \fontsize{9pt}{9pt}\selectfont\textbf{Type} & \fontsize{9pt}{9pt}\selectfont\textbf{Remarks} \\
    \midrule

    \fontfamily{ptm}\small\selectfont $p_i$ & \fontfamily{ptm}\small\selectfont The polarity of page $i$, which is denoted as follows: ``r'' for anti-, ``b'' for pro- and ``g'' for neutral\footnotemark & \fontfamily{ptm}\small\selectfont Categorical & \multirow{4}{\hsize}{\fontfamily{ptm}\small\selectfont In our ML models (i.e., LR, SVR and RFR), we use one-hot encoding to convert categorical features into a binary vector format.} \\ \cmidrule{1-3}
    \fontfamily{ptm}\small\selectfont $c_i$ & \fontfamily{ptm}\small\selectfont The Infomap community assignment of page $i$, encoded with integers ranging from 1 to 172 (i.e., a total of 172 distinct communities were detected) & \fontfamily{ptm}\small\selectfont Categorical & \\ \cmidrule{1-3} 
    \fontfamily{ptm}\small\selectfont $\textit{W-BT}_i$ & \fontfamily{ptm}\small\selectfont The within-group bow-tie role of page $i$, which can be SCC, IN, OUT, and NA & \fontfamily{ptm}\small\selectfont Categorical & \\ 
    \fontfamily{ptm}\small\selectfont $\textit{A-BT}_i$ & \fontfamily{ptm}\small\selectfont The across-group bow-tie role of page $i$, which can be SCC, IN, OUT, and NA & \fontfamily{ptm}\small\selectfont Categorical & \\ 
    \midrule
    
    \fontfamily{ptm}\small\selectfont $f_i$ & \fontfamily{ptm}\small\selectfont The fan count of page $i$ & \fontfamily{ptm}\small\selectfont Numeric & \fontfamily{ptm}\small\selectfont Log transformation implemented to improve our ML model performance\\
    \midrule

    \fontfamily{ptm}\small\selectfont $k_i^{in} := \sum_{j \in \{r, b, g\}} A_{ji}$\footnotemark & \fontfamily{ptm}\small\selectfont The weighted in-degree of page $i$ (recalling that edge weight is obtained by the product of fan counts at both ends), indicating the accumulated strength of recommendations made to page $i$ & \fontfamily{ptm}\small\selectfont Numeric & \multirow{4}{\hsize}{\fontfamily{ptm}\small\selectfont We find that neutral pages seldom reciprocate recommendations with both anti- and pro-vaccination pages. Instead, if neutral pages interact (i.e., recommend or are recommended) with non-neutral counterparts, a clear one-sided leaning emerges. Additionally, anti- (pro-) pages predominantly engage with similar-minded or neutral pages, rarely with pro- (anti-) ones. These trends are visually shown in Supplementary Material. Consequently, we do not delineate features indicating the proportion of inter vaccination group page interactions with specific vaccination groups, and we solely specify the features indicating the proportion of within vaccination group page interactions.}\\

    \fontfamily{ptm}\small\selectfont $k_i^{out} := \sum_{j \in \{r, b, g\}} A_{ij}$ & \fontfamily{ptm}\small\selectfont The weighted out-degree of page $i$, indicating the accumulated strength of recommendations made by page $i$ & \fontfamily{ptm}\small\selectfont Numeric \\[30pt] \cmidrule{1-3}

    \fontfamily{ptm}\small\selectfont $\textit{k-PS}_i^{in} := \frac{\sum_{j \in \{r, b, g\}} \mathbbm{1}_{p_i = p_j} A_{ji}}{\sum_{j \in \{r, b, g\}} A_{ji}} = \frac{\sum_{j \in \{r, b, g\}} \mathbbm{1}_{p_i = p_j} A_{ji}}{k_i^{in}}$ 
    & \fontfamily{ptm}\small\selectfont The proportion of the recommendation strength from pages with the same polarity as page $i$, out of all the accumulated recommendations made to page $i$ & \fontfamily{ptm}\small\selectfont Numeric & \\ 
    \fontfamily{ptm}\small\selectfont $\textit{k-PS}_i^{out} := \frac{\sum_{j \in \{r, b, g\}} \mathbbm{1}_{p_i = p_j} A_{ij}}{\sum_{j \in \{r, b, g\}} A_{ij}} = \frac{\sum_{j \in \{r, b, g\}} \mathbbm{1}_{p_i = p_j} A_{ij}}{k_i^{out}}$ & \fontfamily{ptm}\small\selectfont The proportion of the recommendation strength to pages with the same polarity as page $i$, out of all the accumulated recommendations made from page $i$ & \fontfamily{ptm}\small\selectfont Numeric & \\[50pt] \midrule

    \fontfamily{ptm}\small\selectfont $\textit{k-CS}_i^{in} := \frac{\sum_{j \in \{r, b, g\}} \mathbbm{1}_{c_i = c_j} A_{ji}}{\sum_{j \in \{r, b, g\}} A_{ji}} = \frac{\sum_{j \in \{r, b, g\}} \mathbbm{1}_{c_i = c_j} A_{ji}}{k_i^{in}}$ & \fontfamily{ptm}\small\selectfont The proportion of the recommendation strength from pages with the same Infomap community assignment as page $i$, out of all the accumulated recommendations made to page $i$ & \fontfamily{ptm}\small\selectfont Numeric \\

    \fontfamily{ptm}\small\selectfont $\textit{k-CS}_i^{out} := \frac{\sum_{j \in \{r, b, g\}} \mathbbm{1}_{c_i = c_j} A_{ij}}{\sum_{j \in \{r, b, g\}} A_{ij}} = \frac{\sum_{j \in \{r, b, g\}} \mathbbm{1}_{c_i = c_j} A_{ij}}{k_i^{out}}$ & \fontfamily{ptm}\small\selectfont The proportion of the recommendation strength to pages with the same Infomap community assignment as page $i$, out of all the accumulated recommendations made from page $i$ & \fontfamily{ptm}\small\selectfont Numeric \\
    \midrule 

    \fontfamily{ptm}\small\selectfont $\textit{PageRank}_i$ & \fontfamily{ptm}\small\selectfont The pagerank of page $i$ & \fontfamily{ptm}\small\selectfont Numeric & \multirow{2}{\hsize}{\fontfamily{ptm}\small\selectfont Commonly employed features in research concerning the dynamics of online vaccination-related opinions (e.g., pagerank: \cite{Herrera-Peco-2021, Villegas-2022, Beres-2023}; betweenness: \cite{Herrera-Peco-2021, Olszowski-2022, Durmaz-2022})} \\[12pt]
    \cmidrule{1-3}
    \fontfamily{ptm}\small\selectfont $\textit{Betweenness}_i$ & \fontfamily{ptm}\small\selectfont The betweenness of page $i$ & \fontfamily{ptm}\small\selectfont Numeric & \\[15pt]
    
    \bottomrule
    \end{NiceTabular}
    
    \vspace{10pt}
    \caption{\textbf{Extracted Features from February 2019 Online Recommendation-based Network Snapshot.} We study features of pro- and anti-vaccination pages in order to predict their fan count variations (i.e., $i = r, b$).}	
    \label{tab:feature}
\end{table}

\footnotetext[5]{Recall that to distinguish pages from anti-vaccination, pro-vaccination, and neutral groups, we have used a colour scheme of red, blue, and green. Here we abbreviate them as ``r'', ``b'', and ``g'', respectively.}
\footnotetext{As shown in the beginning of \hyperref[sec:method]{Section 3}, we define $A_{ji} = w > 0$ if there is an edge from node $j$ to $i$ with weight $w$ (i.e., page $j$ recommends page $i$ with recommendation strength $w$) and $A_{j,i} = 0$ otherwise.}

\begin{table}[ht!]
    \begin{subtable}[h]{\textwidth}
        \fontfamily{ptm}\fontsize{6pt}{6pt}\selectfont
	\centering
        \begingroup
        \setlength{\tabcolsep}{1.3pt} 
        \renewcommand{\arraystretch}{2.5} 
	\begin{NiceTabular}{c|ccc|ccccccccccccccccccc}%
            \toprule
            \multicolumn{1}{c}{} & \multicolumn{3}{c}{\multirow{2}{*}{\thead{\textbf{Logistic Regression (LR)} \\ \\(Expanding: 1;\\ Non-expanding: 0)}}} & \multicolumn{9}{c}{\textbf{Support Vector Regression (SVR)}} & \multicolumn{9}{c}{\textbf{Random Forest Regression (RFR)}}\\
            \cmidrule(rl){5-13} \cmidrule(rl){14-22}
            \multicolumn{1}{c}{} & \multicolumn{3}{c}{} & \multicolumn{3}{c}{\textbf{\thead{All \\ anti- \& pro- pages}}} & \multicolumn{3}{c}{\textbf{\thead{Expanding \\ anti- \& pro- pages}}} & \multicolumn{3}{c}{\textbf{\thead{Non-expanding \\ anti- \& pro- pages}}} & \multicolumn{3}{c}{\textbf{\thead{All \\ anti- \& pro- pages}}} & \multicolumn{3}{c}{\textbf{\thead{Expanding \\ anti- \& pro- pages}}} & \multicolumn{3}{c}{\textbf{\thead{Non-expanding \\ anti- \& pro- pages}}}\\
            \cmidrule(rl){2-4} \cmidrule(rl){5-7} \cmidrule(rl){8-10} \cmidrule(rl){11-13} \cmidrule(rl){14-16} \cmidrule(rl){17-19} \cmidrule(rl){20-22}
            \multicolumn{1}{c}{} & \multicolumn{1}{c}{\textbf{Accuracy}} & \multicolumn{1}{c}{\textbf{Sensitivity}} & \multicolumn{1}{c}{\textbf{Specificity}} & \multicolumn{1}{c}{$\mathbf{\text{R}^2}$} & \multicolumn{1}{c}{\textbf{MAE}} &\multicolumn{1}{c}{\textbf{RMSE}} & \multicolumn{1}{c}{$\mathbf{\text{R}^2}$} & \multicolumn{1}{c}{\textbf{MAE}} &\multicolumn{1}{c}{\textbf{RMSE}} & \multicolumn{1}{c}{$\mathbf{\text{R}^2}$} & \multicolumn{1}{c}{\textbf{MAE}} &\multicolumn{1}{c}{\textbf{RMSE}} & \multicolumn{1}{c}{$\mathbf{\text{R}^2}$} & \multicolumn{1}{c}{\textbf{MAE}} &\multicolumn{1}{c}{\textbf{RMSE}} & \multicolumn{1}{c}{$\mathbf{\text{R}^2}$} & \multicolumn{1}{c}{\textbf{MAE}} &\multicolumn{1}{c}{\textbf{RMSE}} & \multicolumn{1}{c}{$\mathbf{\text{R}^2}$} & \multicolumn{1}{c}{\textbf{MAE}} &\multicolumn{1}{c}{\textbf{RMSE}} &\\
            \midrule
            \textbf{All features} & 0.655 & 0.839 & 0.153 & -0.02 & 978.55 & 2854.72 & 0.11 & 1163.30 & 2884.36 & -0.05 & 317.26 & 1088.53 & -0.02 & 1084.06 & 2792.56 & 0.17 & 1127.20 & 2521.70 & -0.18 & 352.83 & 1067.11 \\
            \textbf{Baseline} & 0.517 & 0.490 & 0.525 & -0.02 & 1383.30 & 2808.18 & -0.02 & 1697.50 & 3062.42 & -1.11 & 554.28 & 1048.91 & -0.02 & 1383.30 & 2808.18 & -0.02 & 1697.50 & 3062.42 & -1.11 & 554.28 & 1048.91 \\
            \bottomrule
            \end{NiceTabular}
            \endgroup
            \caption{\textbf{Model Performance.} We apply parameter grid search and 5-fold cross-validation on Logistic Regression (LR), Support Vector Regression (SVR) and Random Forest Regression (RFR). 
            We note that the dataset in LR is unbalanced with 73.0\% expanding pages encoded as ``1'', and for this, we adjust the penalty weights for different classes in the cost function. 
            \textbf{Regarding the evaluation metrics}, we optimise LR using accuracy as its metric, and we include sensitivity and specificity as metrics for further check. 
            In our context, ``sensitivity'' (also known as true positive rate) is explained as the proportion of expanding pages that are correctly classified as such, while ``specificity'' (also known as true negative rate) is explained as the proportion of non-expanding pages that are correctly classified as such.
            We select mean absolute error (MAE) as our evaluation metric to optimise the performance of SVR and RFR, given its robustness to outliers in our dataset.
            We also provide $R^2$ and root mean square error (RMSE) as additional metrics for validation.
            \textbf{Regarding the baseline models}, LR's baseline model uses a random classifier in which each class is assigned an equal probability for prediction; SVR and RFR's baseline models use the mean of fan count variations as their predictors.}
		\label{tab:ml1}
	\end{subtable}
	\hfill
	\begin{subtable}[h]{\textwidth}
		\centering
            \fontfamily{ptm}\fontsize{6pt}{6pt}\selectfont
	    \centering
            \begingroup
            \setlength{\tabcolsep}{10pt} 
            \renewcommand{\arraystretch}{1.5} 
		\begin{NiceTabular}{Y{0.01}c|ccY{0.08}Y{0.08}Y{0.08}Y{0.08}|Y{0.15}Y{0.15}}%
                \toprule
                \multicolumn{2}{c}{} & \multicolumn{2}{c}{\textbf{\thead{All \\ anti- \& pro- pages}}} & \multicolumn{2}{c}{\textbf{\thead{Expanding \\ anti- \& pro- pages}}} & \multicolumn{2}{c}{\textbf{\thead{Non-expanding \\ anti- \& pro- pages}}} & \multicolumn{2}{c}{\thead{Sequential Forward Floating Selection (SFFS) \\ - Using expanding anti- \& pro- pages\\ $\text{Times chosen}$}}\\
                \cmidrule(rl){3-4} \cmidrule(rl){5-6} \cmidrule(rl){7-8} \cmidrule(rl){9-10}
                \multicolumn{2}{c}{} & \multicolumn{1}{c}{\textbf{CC}} & \multicolumn{1}{c}{\textbf{Avg. MI}} & \multicolumn{1}{c}{\textbf{CC}} & \multicolumn{1}{c}{\textbf{Avg. MI}} & \multicolumn{1}{c}{\textbf{CC}} & \multicolumn{1}{c}{\textbf{Avg. MI}} & \multicolumn{1}{c}{\textbf{SVR}} & \multicolumn{1}{c}{\textbf{RFR}}\\
                \midrule
                \multirow{4}{*}{\rotatebox[origin=c]{90}{\textbf{Categorical}}} & $p_i$ & -- & 0.0011 & -- & 0. & -- & 0.0714 & 9 & 48\\
                & $c_i$ & -- & 0.0403 & -- & 0.0002 & -- & 0.1872 & 1 & 28\\
                & \cellcolor[gray]{0.9}$\textit{W-BT}_i$ & \cellcolor[gray]{0.9}-- & \cellcolor[gray]{0.9}0.0131 & \cellcolor[gray]{0.9}-- & \cellcolor[gray]{0.9}0.0001 & \cellcolor[gray]{0.9}-- & \cellcolor[gray]{0.9}0.0788 & \cellcolor[gray]{0.9}34 & \cellcolor[gray]{0.9}21\\
                & \cellcolor[gray]{0.9}$\textit{A-BT}_i$ & \cellcolor[gray]{0.9}-- & \cellcolor[gray]{0.9}0.0345 & \cellcolor[gray]{0.9}-- & \cellcolor[gray]{0.9}0.0424 & \cellcolor[gray]{0.9}-- & \cellcolor[gray]{0.9}0.0179 & \cellcolor[gray]{0.9}36 & \cellcolor[gray]{0.9}36 \\
                \midrule
                \multirow{9}{*}{\rotatebox[origin=c]{90}{\textbf{Numeric}}} & $f_i$ & 0.379 & 0.3331 & 0.485 & 0.4277 & -0.372 & 0.2884 & 50 & 50\\
                 & $k_i^{in}$ & 0.406 & 0.1013 & 0.416 & 0.1501 & -0.036 & 0.0770 & 50 & 41\\
                 & $k_i^{out}$ & 0.254 & 0.0199 & 0.263 & 0.0360 & -0.267 & 0.0012 & 50 & 42\\
                 & $\textit{k-PS}_i^{in}$ & -0.039 & 0.0458 & -0.107 & 0.0636 & 0.237 & 0.0033 & 40 & 29\\
                 & $\textit{k-PS}_i^{out}$ & -0.078 & 0.0008 & -0.137 & 0. & 0.145 & 0.0432 & 49 & 50\\
                 & $\textit{k-CS}_i^{in}$ & 0.091 & 0.0088 & 0.053 & 0.0037 & 0.275 & 0.0117 & 49 & 39\\
                 & $\textit{k-CS}_i^{out}$ & 0.052 & 0.0228 & 0.044 & 0.0170 & 0.074 & 0.0052 & 39 & 37\\
                 & $\textit{PageRank}_i$ & 0.252 & 0.0958 & 0.258 & 0.1203 & -0.032 & 0.5394 & 50 & 40\\
                 & $\textit{Betweenness}_i$ & 0.142 & 0.0170 & 0.127 & 0.0170 & 0.032 & 0.0008 & 43 & 38\\
            \bottomrule
            \end{NiceTabular}
            \endgroup
		\caption{\textbf{Feature Comparison.} Due to the stochastic nature of both Mutual Information (MI) and Sequential Forward Floating Selection (SFFS), we conduct 50 runs for both methods and present the average MI value for each feature and the selected feature frequencies, respectively.
        Furthermore, the Sequential Forward Floating Selection (SFFS) model identifies the optimal feature subset comprising 10 features to ensure that at least one categorical feature is selected in each run (i.e., for categorical feature comparison).}
		\label{tab:ml2}
	\end{subtable}
	\hfill
	\label{tab:ml}
        \caption{\textbf{Results about supervised machine learning.}}
\end{table}

To investigate whether our supervised ML models are more feasible in predicting fan count fluctuations for expanding anti- \& pro- pages (i.e., with an increased fan count) over non-expanding ones (i.e., with an unchanged or decreased fan count), we first use all features in Table \ref{tab:feature} in Logistic Regression (LR) to predict page status as either expansion or non-expansion.
Furthermore, with the same goal in mind, we employ two supervised machine learning models: Support Vector Machine (SVR) and Random Forest Regression (RFR) to investigate the precise fan count variations within three categories of pages: all anti- \& pro- pages, anti- \& pro- pages that experienced expansion between February and October 2019, and anti- \& pro- pages that remained non-expanded. 
Results are indicated in Table \ref{tab:ml1}.
Our findings show that Logistic Regression moderately distinguishes between page expansion and non-expansion (Accuracy: 0.655), with a stronger capability to identify expanding pages over non-expanding ones (Sensitivity: 0.839; Specificity: 0.153). This aligns with our initial expectation that recommendations can correlate more with fan size increase than decrease. 
Additionally, these results are supported by the notable performance improvements of SVR and RFR when specifying predictions exclusively for expanding pages, compared with predictions for non-expanding pages.
For example, the $R^2$ values of SVR and RFR on expanding pages are $0.11$ and $0.17$, which are much higher than those for non-expanding pages, $-0.05$ and $-0.18$, and those for all anti- and pro- pages, $-0.02$ and $-0.02$.

To evaluate the significance of each feature in our predictions above, we employ the correlation coefficient (CC) to measure linear dependencies between numeric features and fan count variations, again across three page categories (i.e., all anti- \& pro- pages, expanding anti- \& pro- pages, and non-expanding anti- \& pro- pages). We also apply mutual information (MI) to capture non-linear dependencies among all numeric and categorical features and fan count variations within these three page categories.  
Moreover, we employ Sequential Forward Floating Selection (SFFS) \cite{Chandrashekar-2014}, a widely used feature selection method, to identify the optimal feature subset yielding the best performance in SVR and RFR models, exclusively for expanding anti- \& pro- pages only. 
This decision is based on the better interpretability and performance of predictions for expanding pages, as previously confirmed.
The results are presented in Table \ref{tab:ml2}.
Our findings indicate that among categorical features, bow-tie relevant features (i.e., $\textit{W-BT}_i$ and $\textit{A-BT}_i$) exhibit relatively strong performance, with $\textit{A-BT}_i$ outperforming $\textit{W-BT}_i$. 
For example, the MI of $\textit{A-BT}_i$ regarding expanding anti- \& pro- pages is notably higher than other categorical features. Also, SFFS reveals that both $\textit{W-BT}_i$ and $\textit{A-BT}_i$ significantly surpass polarity $p_i$ and Infomap community $c_i$ in the SVR model, although this advantage diminishes somewhat in the RFR model. 
Results about numeric features help explain the interpretability of our models. 
Of all numeric features, $f_i$ (i.e., the page fan count in Feb 2019) exhibits the highest significance across all three page categories, as examined through CC, MI, and SFFS.
This observation, coupled with the opposite sign of its CC within expanding and non-expanding pages ($0.485$ vs $-0.372$), may indicate the ``snowball effect'' \cite{snowball}, where the fan count of each page tends to either increasingly grow or decreasingly drop over time (i.e., pages with larger initial fan counts generally experience more substantial changes).
Additionally, the higher significance of $k_i^{in}$ compared to $k_i^{out}$ is reasonable, with the former representing the strength of recommendations directed to each page and the latter indicating recommendations made by each page.

Overall, supervised machine learning predictions of page fan count changes from February to October, based on the February online recommendation network, are generally more feasible and interpretable for expanding pages. 
Features related to bow-tie structure demonstrate potential in helping to predict the fan count variations of expanding pages.


\subsubsection{Agent-based SIR Model}
In our online recommendation-based networks, a recommendation from page A to page B is often triggered when page B shares content that captivates the interest of members of page A. This positive exchange of information may lead to page B acquiring new fans from page A.
Motivated by this, we use a SIR model to simulate information cascades from each page and compare their information influence with fan count variations.

The SIR (Susceptible - Infected - Recovered) model, initially developed for simulating disease spread \cite{Istvan-2017}, has been used for modelling information diffusion in social networks \cite{Cinelli-2021, Cinelli-2020, Kabir-2019, Qiu-2021}. 
In a network-based SIR model, each node can be in any of three states: susceptible (unaware of the circulating information), infectious (aware through creation or reception and willing to spread further), or recovered (aware but no longer transmitting, e.g., due to information obsolescence).
Susceptible nodes can transition to an infectious state through contact with infected neighbours, the probability of which is proportional to the transmission rate $\beta$. Infectious nodes can transition to a recovered state spontaneously, with a probability of $\gamma$. 
A SIR epidemic process is usually initialised with one randomly selected node $i$ being infected and terminated when no infectious nodes remain, simulating the propagation of a singular piece of information \cite{Cinelli-2021}. 

In our dataset, we run the agent-based SIR model on the February 2019 recommendation-based network.
Consider to this end a total of $N$ pieces of information generated during the period from February to October (for comparison with the October 2019 snapshot). 
For each individual piece of information $n \in  \{1, 2, ..., N\}$, the initial page that generates it, denoted as $I_n^0$, is selected randomly from a pool of anti- and pro- pages, following the probability distribution $\mathds{P}^{\text{initialiser}}$\footnote{Note that we prevent the initialisation of information from neutral pages, as their information pieces are not our primary focus, in line with our previously mentioned assumption. That is, $\mathds{P}^{\text{initialiser}}_i:= \mathds{P}(I_n^0 = i) = 0, \text{ if } p_i = g$. As indicated in Table \ref{tab:feature}, $p_i \in \{r,b,g\}$ denotes the polarity of page $i$ for anti-, pro-, neutral, respectively.}.
The set of pages that are impacted by information piece $n$ (i.e., in a recovered state at the end of the SIR dynamics about information $n$) is represented as $I_n$.
The influence of information piece $n$ is defined as the aggregate fan counts of the pages impacted by it, represented by $\sum_{j \in I_n} f_j$. 
Subsequently, the influence of a page $i$ is defined as the accumulated influence of all information pieces originating from page $i$, represented by $\sum_{n = 1}^N\mathbbm{1}_{I_n^0 = i} (\sum_{j \in I_n} f_j)$. 
In order to account for how such information spread relate to the bow-tie components, we further divide the influence of each information piece into two categories: within-group and across-group.
The former accounts for the influence of information piece $n$ initialised from a pro- or an anti- page to pages with the same vaccination stance, and the latter accounts for the influence of information piece $n$ from a pro- or an anti- page to neutral pages.
Note that we disregard the influence of information piece $n$ from a pro- page to an anti- page (or the other way around), as we assume users who follow pro- (anti-) pages may hardly accept information from anti- (pro-) pages and consequently follow them\footnote{This assumption is also supported by the dataset, as evidenced by the sparse direct recommendations between anti- pages and pro- pages indicated in Figure \ref{fig:data}c, and the Infomap community detection result in Figure \ref{fig:obs}.}.
We define the influence measures as follows:
\begin{align*}
    \textit{Info W-Influ}_n = \sum_{j \in I_n} \mathbbm{1}_{p_{I_n^0} = p_j}f_j, \: p_{I_n^0} \in \{r, b\}\\
    \textit{Info A-Influ}_n = \sum_{j \in I_n} \mathbbm{1}_{p_j = g}f_j, \: p_{I_n^0} \in \{r, b\}
\end{align*}
Subsequently, the within-group and across-group influence of a pro- or an anti- page $i$ are defined below:
\begin{align*}
    \textit{Page W-Influ}_i = \sum_{n = 1}^N\mathbbm{1}_{I_n^0 = i} (\sum_{j \in I_n} \mathbbm{1}_{p_i = p_j}f_j), \: I_n^0 \sim \mathds{P}^{\text{initialiser}}\\
    \textit{Page A-Influ}_i = \sum_{n = 1}^N\mathbbm{1}_{I_n^0 = i} (\sum_{j \in I_n} \mathbbm{1}_{p_j = g}f_j), \: I_n^0 \sim \mathds{P}^{\text{initialiser}}
\end{align*} 
Such within-group and across-group influences establish connections with within-group and across-group bow-tie components, respectively. This enables our following analysis, where we explore the potential influence distinctions among various bow-tie components, both within-group and across-group, and investigate whether these distinctions can aid in extracting pages' roles in terms of information flow (i.e., SCC - ``magnifiers'', OUT - ``creators'', and IN - ``listeners'') and predicting variations in fan counts for each page.

\begin{figure}[bpht]
    \centering
    \adjincludegraphics[width=0.9\textwidth,trim={{.01\width} {.32\height} {.01\width} {.22\height}},clip]{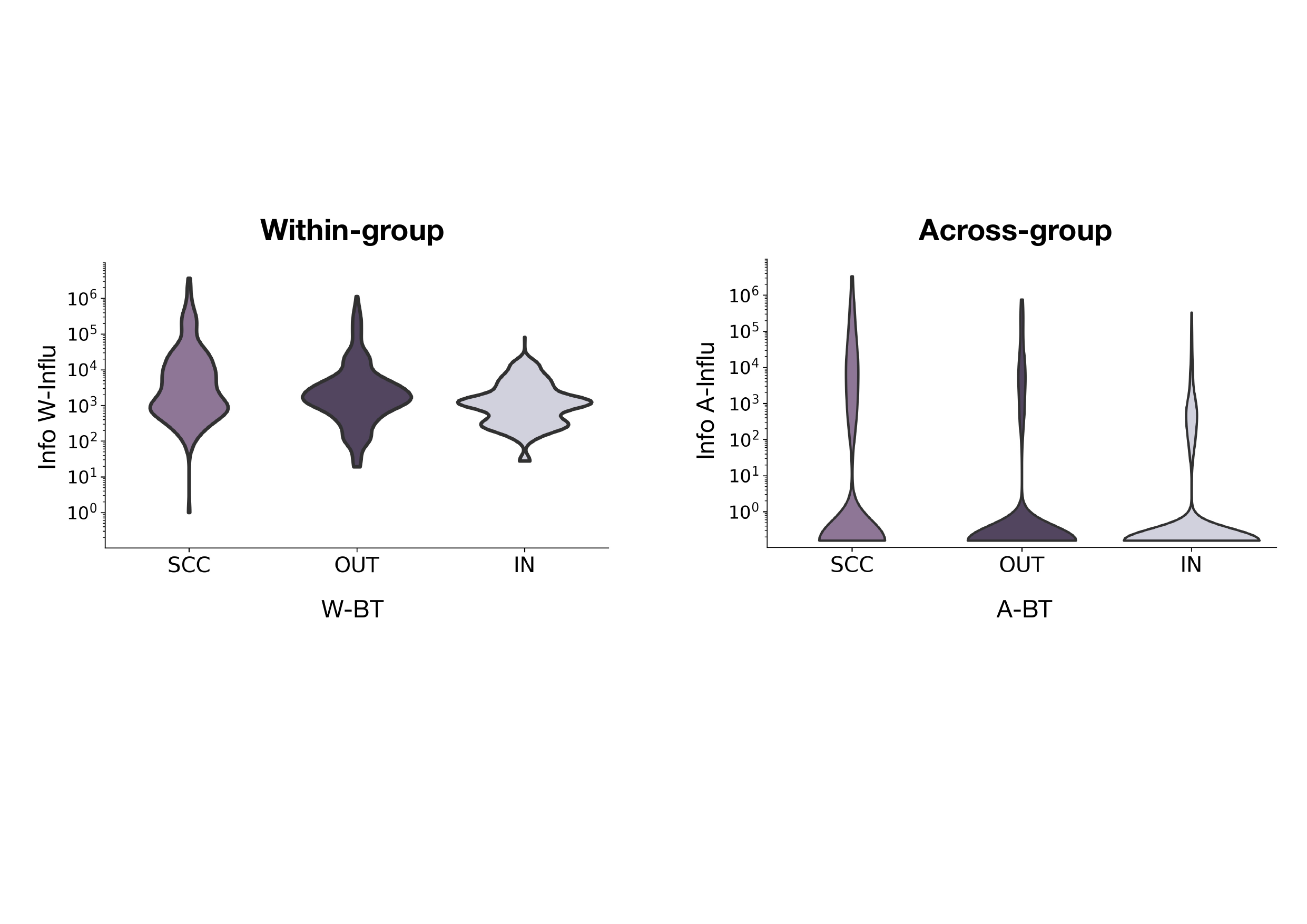}
    \caption{\textbf{Distinctions among the influence of information pieces initialised from different bow-tie components.} 
    Each bow-tie component generates $1000$ information pieces, with each page inside a component holding an equal probability of being the initial source for a single information piece, and the SIR epidemic process for each information piece is set at $\beta = 0.5$ and $\gamma = 0.3$.
    The violin plots depict the distribution of information within-group (across-group) influence, based on initialisation from different within-group (across-group) bow-tie components.
    We observe that the hierarchy of influence, both within-group and across-group, adheres to the following ranking: SCC > OUT > IN.
    This aligns with the roles of these components (i.e., SCC - ``magnifiers'', OUT - ``creators'', and IN - ``listeners'') formed by the bow-tie decomposition.
    Additionally, a large quantity of information pieces remain confined within-group and have limited dissemination across-group, mirroring the real-world dynamics. 
    }
    \label{fig:sir1}
\end{figure}

We first examine disparities in the influence of information pieces (i.e., $\textit{Info W-Influ}_n$ and $\textit{Info A-Influ}_n$) originating from different bow-tie components.
Our results in Figure \ref{fig:sir1} show that the hierarchy of influence, both for within-group and across-group, adheres to the following ranking: SCC > OUT > IN. 
This aligns with the roles of these components (i.e., SCC - ``magnifiers'', OUT - ``creators'', and IN - ``listeners'').
Also, a substantial quantity of information pieces remain confined within-group and have limited dissemination across-group, mirroring the real-world dynamics.

\begin{figure}[bpht]
    \centering
    \adjincludegraphics[width=\textwidth,trim={{.03\width} {.11\height} {.03\width} {0.07\height}},clip]{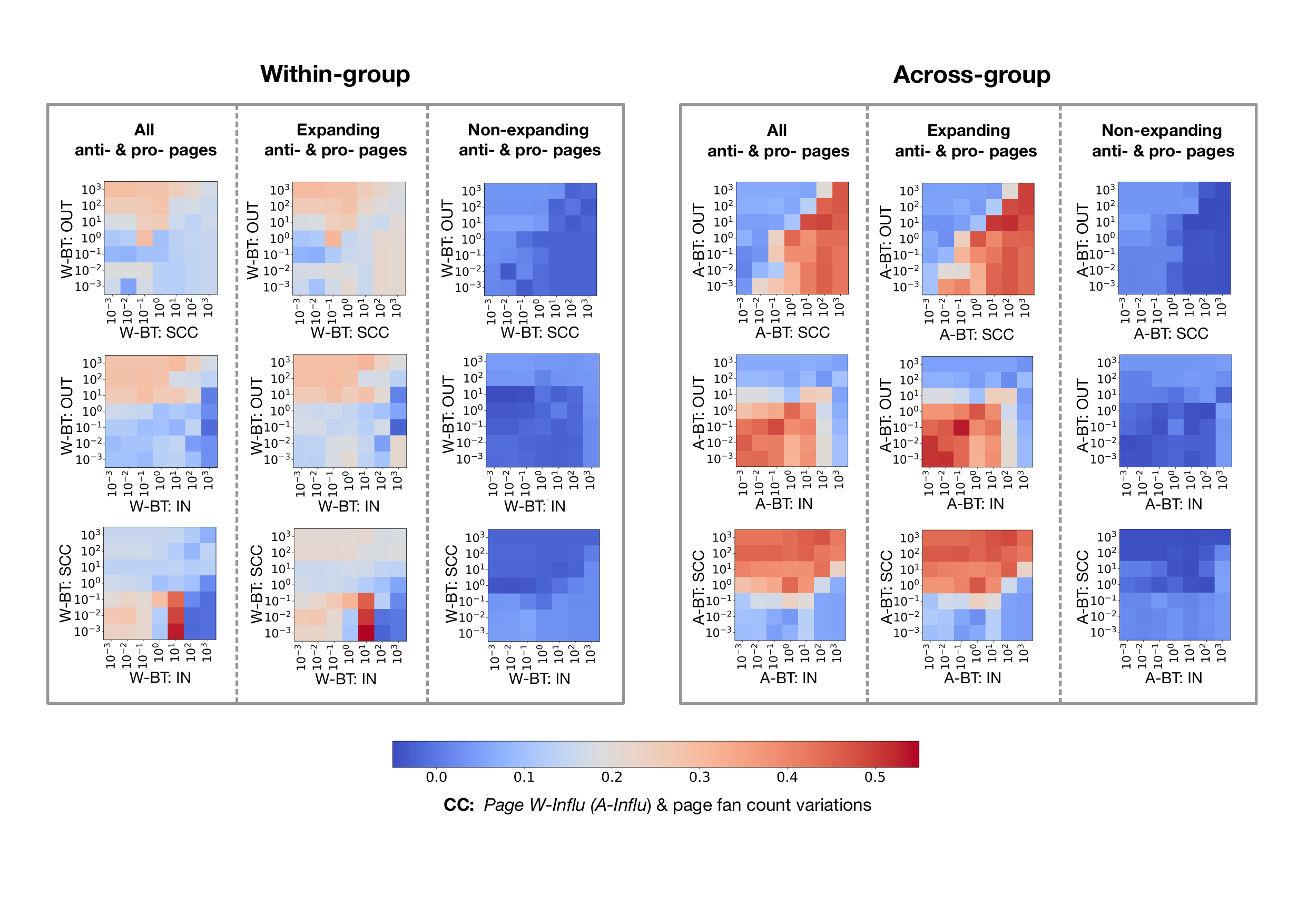}
    \caption{\textbf{Distinctions among the influence of pages in different bow-tie components when varying their probability of generating information pieces.} 
    Each heatmap pixel represents the correlation coefficient (CC) between the page influence and the page fan count fluctuations during the specified SIR epidemic process.
    We generate $N = 3000$ pieces of information for each pixel with $\beta = 0.5$ and $\gamma = 0.3$ (aligning with Figure \ref{fig:sir1}), and customise the probability of generating information $\mathds{P}^{\text{initialiser}}$ for pages in different bow-tie components. 
    Specifically, in each heatmap, the x and y axes represent the information generation probabilities for pages in the respective bow-tie components, divided by the information generation probabilities for pages in other components that are not involved in either axis.
    Our results indicate that both within-group and across-group page influence appear more correlated with fan count variations of expanding pages, in contrast to non-expanding pages.
    The higher CCs of across-group page influence than within-group potentially suggest that the increase in fan counts for anti- and pro- pages may be more strongly influenced by their interactions with neutral pages instead of similar-minded ones.
    In light of these two points, the upper-left red corner in the \textit{W-BT:OUT} and \textit{W-BT:SCC} heatmap illustrates that within-group OUT pages (``creators'') are likely to produce more information pieces that recruit fans of pages sharing the same vaccination stance, compared with SCC pages. 
    This result is also supported by other within-group heatmaps. 
    Conversely, across-group SCC pages (``magnifiers'') tend to generate more information pieces that possibly recruit neutral pages' fans compared with OUT pages. 
    Both across-group and within-group IN pages (``listeners'') exhibit a limited trend in generating information pieces contributing to fan size variations, though surprisingly having a good performance when $x = 10$ in the heatmap related to \textit{W-BT:SCC} and \textit{W-BT:IN}.
    Moreover, the CCs of within-group and across-group page influence can maintain relatively high values compared with other numeric features in Table \ref{tab:ml2}. This demonstrates their high potential in aiding the prediction of fan count increases for anti- and pro- pages.
    }
    \label{fig:sir2}
\end{figure}

Secondly, we explore the disparities among the influence of pages (i.e., $\textit{Page W-Influ}_i$ and $\textit{Page A-Influ}_i$) in different bow-tie components, to investigate whether pages in certain bow-tie components tend to generate more information pieces that contribute to their fan count variations, by adjusting $\mathds{P}^{\text{initialiser}}$.
We use correlation coefficient (CC) as the metric to measure the association between page influence and fan count variations, as $\mathds{P}^{\text{initialiser}}$ varies\footnote{
In Figure \ref{fig:sir2}, each heatmap adjusts page generation probabilities $\mathds{P}^{\text{initialiser}}$ based on x and y axis values and associated bow-tie components ($BT_X$ and $BT_Y$).
For each pixel $(x, y)$ in the heatmap, we generate $N$ information pieces, with each page $i$ having a probability $\mathds{P}^{\text{initialiser}}_i := \mathds{P}(I_n^0 = i) \propto m_i$, scaled to sum up to 1, where $m_i = x, y$ if $i \in BT_X, BT_Y$, and $m_i = 1$ otherwise. In this sense, the x and y axes intuitively represent the information generation probabilities for pages in the respective bow-tie components, divided by the information generation probabilities for pages in other components that are not involved in either axis. Finally, the colour in each heatmap pixel represents the correlation coefficient (CC) between the page influence and the page fan count fluctuations in the specified situation.
}.
Our results are shown in Figure \ref{fig:sir2}.
As we can observe, both within-group and across-group page influence appear more correlated with fan count variations of expanding pages, in contrast to non-expanding pages, aligned with our expectation (i.e., recommendations correlate more with fan size increase than decrease) and our ML results above.
The higher CCs of across-group page influence than within-group potentially suggest that the increase in fan counts for anti- and pro- pages may be more strongly influenced by their interactions with neutral pages instead of similar-minded ones. This again aligns with our previous results in ML models (i.e., $\textit{W-BT}_i$ vs $\textit{A-BT}_i$).
In light of these two points, the skewed CC heatmaps in our results illustrate that within-group OUT pages are likely to produce more information pieces that recruit fans of pages sharing the same vaccination stance, compared with SCC pages. Conversely, across-group SCC pages tend to generate more information pieces that possibly recruit neutral pages' fans compared with OUT pages. 
This finding helps explain a possible scenario, where OUT pages (``creators'') often generate ``innovative'' content that captures the interest of like-minded users (within-group), potentially inspiring content in SCC pages, while SCC pages (``magnifiers'') amplify certain ``mature'' content across-group, targeting neutral pages. 
Both across-group and within-group IN pages (``listeners'') exhibit a limited trend in generating information pieces contributing to fan size variations, though surprisingly having a good performance when $x = 10$ in the heatmap related to \textit{W-BT:SCC} and \textit{W-BT:IN}.
Notably, compared with the CCs between fan count variations of expanding anti- and pro- pages and other numeric features in Table \ref{tab:ml2} (e.g., February fan count $f_i$: $0.485$; $\textit{PageRank}_{i}$: $0.258$), the ones of our within-group page influence can maintain around $0.25$, and across-group page influence around $0.5$, demonstrating high potential in helping predict fan count increase for anti- and pro- pages.

Finally, the results above remain robust across different parameter choices for transmission rate $\beta$ and recovery rate $\gamma$. See details in Supplementary Material. 

\section{Conclusion and Future Work}
\label{sec:discussion}
In this paper, we investigate bow-tie structure of discursive communities (i.e., groups of users sharing common communicative purposes) in temporal online social networks that describe the recommendations between anti-vaccination, pro-vaccination, and neutral pages, with snapshots taken in February and October 2019.

By employing dual interpretations of discursive communities (one as vaccination groups and the other as communities detected by Infomap), we perform bow-tie analyses of recommendation networks within and across vaccination groups. 
Our results indicate different bow-tie structures among various vaccination groups. In both bow-tie analyses, a large number of anti-vaccination pages are assigned OUT bow-tie roles, while a substantial portion of pro-vaccination pages are assigned SCC bow-tie roles. These bow-tie structures exhibit statistical significance and demonstrate stability over the considered timeframe. In contrast, the neutral group displays different bow-tie structures across these two analyses and demonstrates less temporal stability than the pro- and the anti- group.

We then relate these detected bow-tie structure differences to opinion dynamics, investigating their potential to predict changes in page fan counts from February to October using the February network. 
We implement both supervised machine learning models involving a variety of features, and mechanistic models on information cascades focusing on explainability, with these two approaches complementing and validating each other's results. 
All our models are more adept at predicting page expansion (i.e., an increase in fan count) over non-expansion (aligning with our expectations), and bow-tie structure features exhibit promise in enhancing the prediction for expanding pages.
Notably, such promise is indicated both in the performance of our models (e.g., bow-tie features $\textit{W-BT}_i$ and $\textit{A-BT}_i$ significantly surpass polarity $p_i$ and Infomap community $c_i$ in our SFFS-SVR machine learning model), and in the high interpretability of our agent-based models. For example, in our mechanistic simulation model, within-group OUT bow-tie pages - ``creators'' are shown to produce more information pieces that possibly recruit fans of pages sharing the same vaccination stance, while across-group SCC bow-tie pages - ``magnifiers'' tend to generate more information pieces that possibly recruit neutral pages' fans.

\subsection{Future Work}
There are a number of interesting directions to explore in future work.
Firstly, the large OUT and SCC bow-tie components detected in the anti- and pro-vaccination groups, suggest distinct advantages held by the anti- and pro-vaccination groups: the former exhibits a strong commitment to generating information, while the latter possesses a strong capability for information dissemination. Based on our findings, which indicate that bow-tie structures can aid in predicting increases in their fan counts at the page level, it would be interesting to develop a bow-tie-based model for a longer timescale fan count prediction of anti- and pro-vaccination pages spanning decades, contrasting it with Johnson et al.'s model \cite{Johnson-2020}, which disregards page recommendations in the dataset. 

Secondly, for this bow-tie-based long-term fan count prediction model mentioned above, it would be interesting to incorporate the temporality of bow-tie structures, in light of the stronger temporal stability observed in the anti- and pro-vaccination groups compared to the neutral group in our dataset.
More broadly, this framework can be expanded into a generative model that can be used to produce synthetic networks with different levels of bow-tie role adherence (e.g., SCC - ``magnifiers'', OUT - ``creators'', and IN - ``listeners''). These synthetic networks can serve as benchmarks and aids for inferring the structure of empirical networks of interest. 
We emphasise that our interest in role structure in information flow is not uncommon \cite{Elliott-2020, Beguerisse-2014, Bovet-2022, Farzam-2023} (e.g., Beguerisse-Díaz et al. \cite{Beguerisse-2014} capture five different roles in Twitter users, including ``listeners'', ``diversified listeners'', ``references'', ``engaged leaders'' and ``mediators''), therefore developing such models can be important to understand the behavioural ecology in online social networks. 
The reference \cite{Bazzi-2020} can be a useful starting point for constructing the modelling framework. 

Thirdly, the approaches considered in this paper are general and can be applied to a range of social networks.   
For instance, we can use bow-tie structure to investigate potential distinctions between misinformation and scientific information dissemination. 
A recent paper \cite{Castioni-2022} highlights that a minority of accounts are responsible for the majority of the misinformation circulating on Twitter, a pattern highly pertinent to our bow-tie structure analysis.
Bow-tie structure may also help explain the phenomenon of infodemic in misinformation circulation, which describes the situation where exposure to an abundance of information undermines people's ability to discern disinformation, thus facilitating its dissemination \cite{Mattei-2022, Briand-2021}. 
More broadly, bow-tie structure analysis can yield insights about the structure and evolution of social networks, which we hope will be a helpful addition to those designing intervention efforts that aim to mitigate misinformation in social networks.

Fourthly, when examining the relationship between bow-tie structure and page fan count variations, we perform mechanistic simulations on information cascades. For our purpose, we use the agent-based SIR model in order to incorporate bow-tie-related factors. There are several other models that one could use. These include information dissemination models that incorporate additional factors that may be relevant to our research questions \cite{Arruda-2022, Picu-2012, Zhang-2023}. For instance, \cite{Arruda-2022} introduces a ``forgetting mechanism'' affecting rumour lifespan and \cite{Picu-2012} incorporates ``delays'' in information spreading.

In this paper, we study opinion dynamics through the lens of fan count change.
The literature on opinion dynamics is broad and diverse, with models ranging from independent cascade to threshold models \cite{grabish-rusinowska2010, KempeKT05}, widely studied in the social and computational sciences \cite{ZhaoEcommerce, EasleyKleinberg}. Usually, opinion diffusion models analyse the network dynamics at the individual node level, e.g., with agents changing their minds based on the majority of their influencers. Our approach is different, as we look at nodes with expanding volumes - our fan sizes - based on the polarity of selected influencers and their own size. This suggests a novel framework for understanding opinion dynamics, whose study is interesting in its own sake.

\vspace{.5cm}

\noindent \textbf{Data Accessibility}
The dataset in this paper was provided by Johnson et al. \cite{Johnson-2020} and Illari et al. \cite{Illari-2022} spanning different time periods. The former contains two snapshots in February 2019 and October 2019 (before the COVID-19), which we study in our main paper. The latter contains another two snapshots in November 2019 and December 2020 (at the initial stage of the COVID-19), which we study in Supplementary Material. Both of them are openly available and documented in two papers separately of different formats (PDF \& Excel), thus requiring intensive preprocessing.

To make it easier for other researchers to use this dataset, we reorganise both versions of this dataset in gpickle format (can be directly imported as attributed networks using Python) and in CSV format (for general use of the dataset) in \href{https://github.com/YuetingH/BT_Vaccination_Views}{GitHub} and have been archived within the \href{https://zenodo.org/records/10513913}{Zenodo repository}. We also document all the programming code used in this paper there. 

\vspace{.5cm}

\noindent \textbf{Funding}
Yueting Han acknowledges support from EPSRC grant no. EP/S022244/1 through the MathSys CDT, University of Warwick. Paolo Turrini acknowledges the support of the Leverhulme Trust for the Research Grant RPG-2023-050 titled ``Promoting Social Good Using Social Network''.


\bibliographystyle{unsrt}  
\bibliography{references}  

\clearpage

\section*{Supplementary Material}
\subsection*{A. Methodology: Infomap Community Detection in Across-group Bow-tie Structure}

The across-group bow-tie decomposition in our paper uses community detection to partition the entire graph into subgraphs and apply bow-tie decomposition to each of these subgraphs. 
Here we explain why we choose Infomap over modularity maximisation as our community detection method for our specific problem, and examine the stochasticity of Infomap in our dataset. 

\paragraph{Infomap vs Modularity Maximisation}

\begin{figure}[bpht]
  \centering
  \adjincludegraphics[width=0.6\textwidth,trim={{.05\width} {.28\height} {.08\width} {.25\height}},clip]{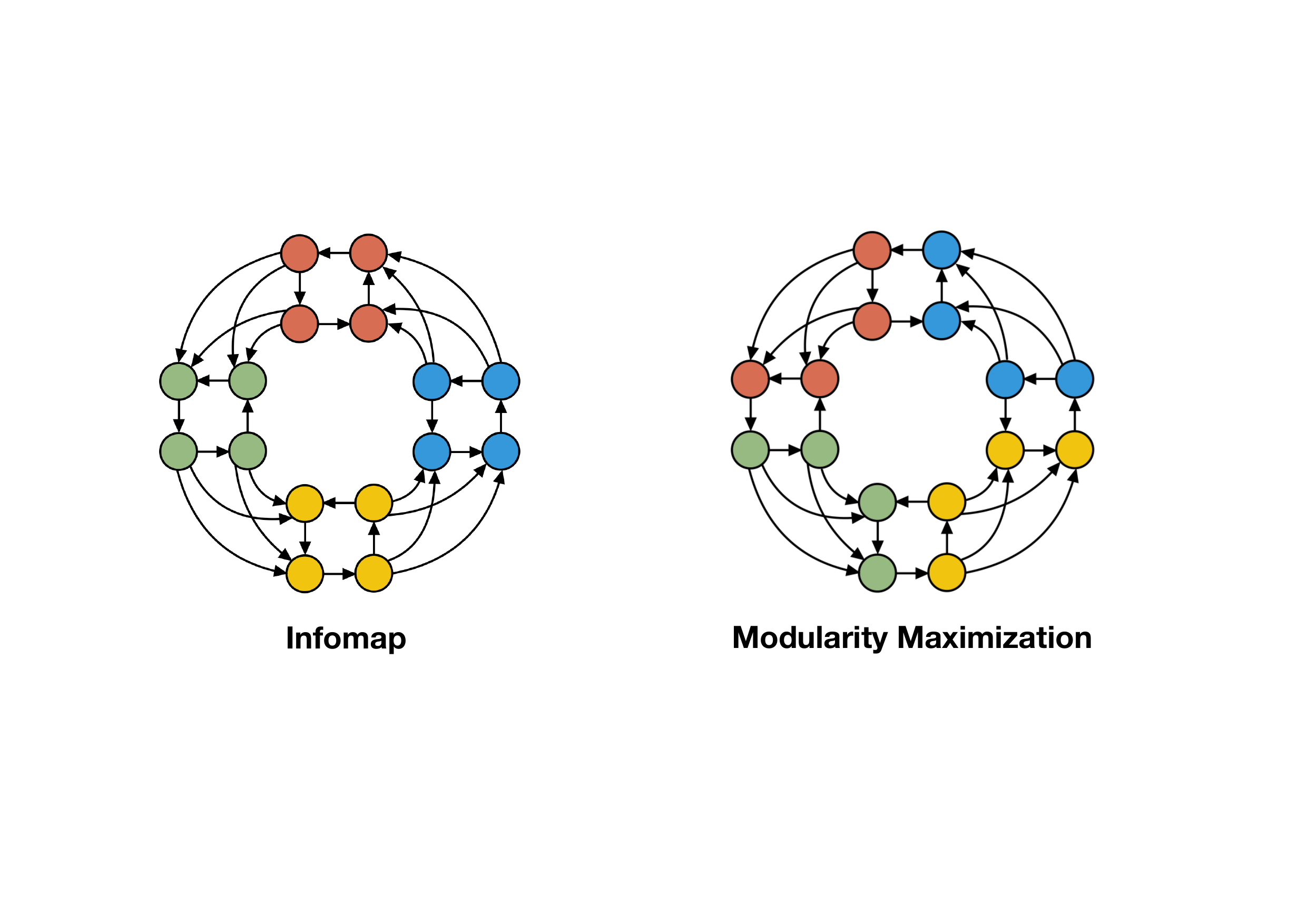}
  \caption{\textbf{Community Detection Method Comparison.} 
  Infomap and Modularity Maximisation are implemented on a directed unweighted graph. 
  Colours illustrate community assignment results. 
  By observation, Infomap appears to gather mutually reachable groups whereas modularity maximisation tends to detect densely intertwined components.}
  \label{fig:comm}
\end{figure}

In comparison to traditional modularity maximisation \cite{comm_newman}, we chose to use Infomap \cite{comm} due to its ability to capture directed flow-based network features. This aligns well with our dataset that represents directed and dynamic recommendations among Facebook pages. 
Figure \ref{fig:comm} further demonstrates the difference between Infomap and modularity maximisation using an example network. 

\begin{figure}[bpht]
\centering
\adjincludegraphics[width=0.7\textwidth,trim={{.03\width} {.28\height} {.03\width} {.17\height}},clip]{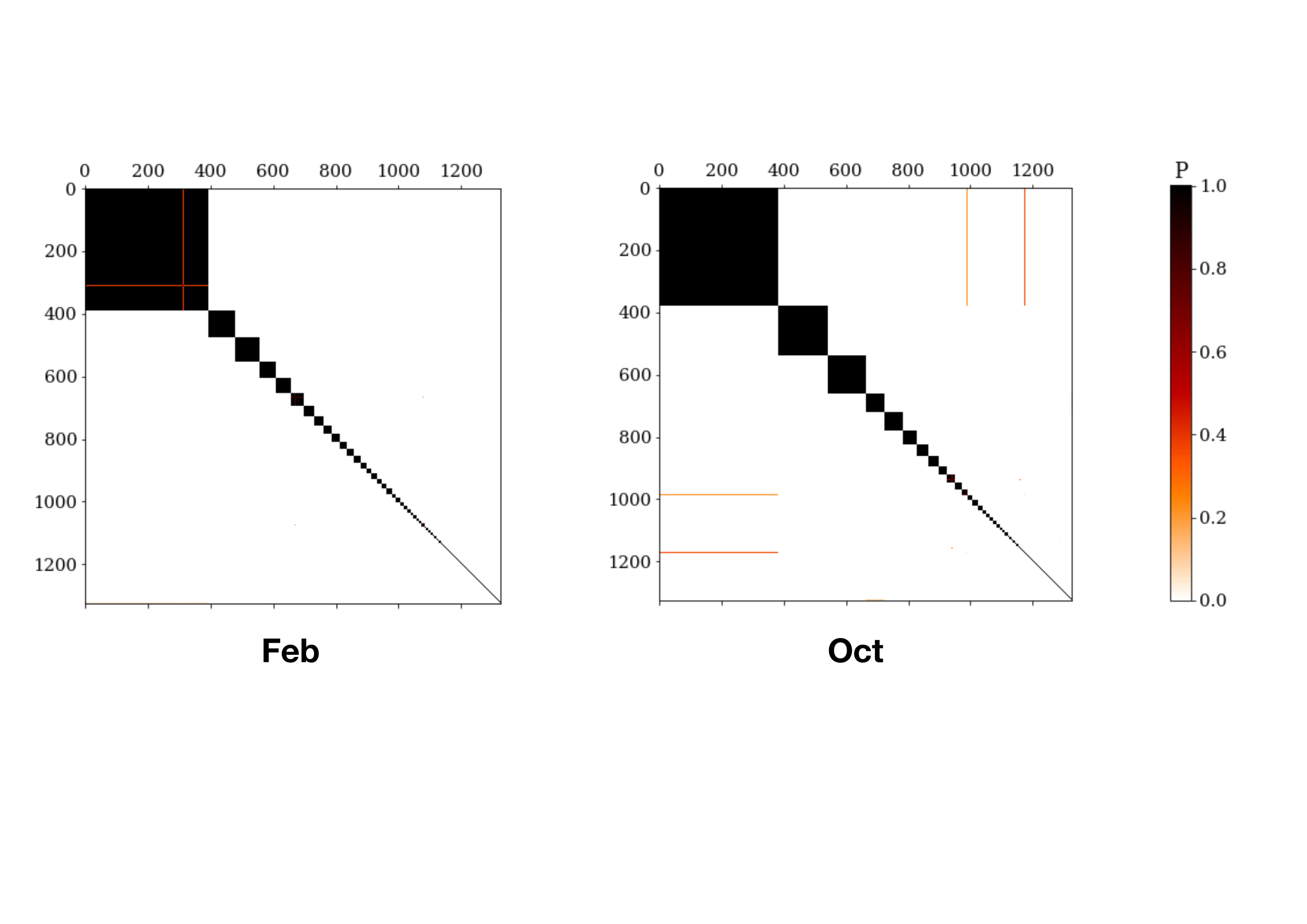}
\caption{\textbf{Stochasticity of Infomap in our dataset.} For each network snapshot in February and October 2019, we use an association matrix to visualise the stochasticity of the Infomap implementation we use.
The colour in each pixel represents the proportion of two nodes assigned to the same community in 50 runs of the algorithmic heuristic, denoted by $P$.
Pixels in red, corresponding to middle values of $P$, indicate unstable community assignments.
Note that nodes have been rearranged to list sets of communities detected in the first run in descending order of their sizes.
Overall, the algorithmic heuristic used demonstrates strong robustness in our dataset, with the majority of pixels appearing either white or black.
}
\label{fig:stochasticity}
\end{figure}

\paragraph{Infomap Stochasticity}
As depicted in Figure \ref{fig:stochasticity}, partitions identified with Infomap implementation used are overall robust to algorithmic stochasticity in our dataset, primarily due to its deterministic greedy search for the main result, with randomness involved only in its refinement process \cite{comm}.
The nodes with unstable community assignments (i.e., with red pixels) in February and October are [``a\_00586'', ``n\_00555'', ``n\_00587''] and [``n\_00230'', ``a\_00586'', ``n\_00555'', ``n\_00587'', ``n\_00605'']\footnote{Nodes (Pages) with IDs starting from ``a''/``n''/``p'' are anti-/neutral/pro-vaccination view supporters, respectively.}. 
Most of them represent pages with a neutral vaccination stance. There are also some overlapping nodes in both sets, suggesting the connection between two network snapshots reflected by Infomap.

\subsection*{B. Bow-tie Detection in the Recommendation-based Networks}

\subsubsection*{B.1. Statistical Significance}

\paragraph{Background.}
Every directed graph can be represented by a (recursive) bow-tie structure, as mentioned in our main paper. How statistically significant such a representation is becomes a key problem \cite{Mattei-2022, Elliott-2020, bt-evaluation}. 
One could argue that the presence of (recursive) bow-tie structure is not significant, as it may be primarily attributed to the degree distribution of a graph. For example, a node with a low in-degree and high out-degree is likely to be distributed to OUT component, while a node with the opposite features can naturally belong to IN component.
This motivates the use of random graph models to assess the extent to which the emergence of (recursive) bow-tie structure is due to a random organisation of edges, given a fixed degree sequence. 
Our approach for evaluating the statistical significance of (recursive) bow-tie structure retains the in- and out-degree distribution preserves each page's incentive to proactively recommend other pages and their potential to be directly recommended, regardless of the particular source or target of these recommendations, respectively.

\paragraph{Algorithm.}
We assess the statistical significance of recursive bow-tie structure by individually examining the statistical significance of bow-tie structure within each of its partitioned subgraphs\footnote{Recall that partitioning the entire network into subgraphs is a necessary step when performing recursive bow-tie decomposition.}. 
We therefore only explain the procedure for evaluating the statistical significance of bow-tie structure below.

Firstly, a benchmark model is chosen to generate directed random graphs based on the observed graph, where both in- and out-degree sequences are conserved. In our paper, we implement the commonly used directed configuration model from Newman et al. \cite{dcm}, which satisfies this key assumption.

Based on the selected benchmark model, the procedure to assess the statistical significance of bow-tie structure is listed in \hyperref[algorithm:sta]{Algorithm 1}. 
The key idea is that, for each bow-tie component within the observed graph present in our dataset, we investigate how large/small it is compared with the ones in generated random graphs. If the component is too large or too small, it suggests its statistical significance. To quantify this, we introduce a metric referred to as ``rank R'', which represents the percentage of random graphs in which the corresponding bow-tie component is smaller than the observed graph. A high or low R value indicates the statistical significance of the corresponding bow-tie component.

\begin{algorithm}[bpht]
  \label{algorithm:sta}
  \caption{Statistical Significance of Bow-tie Structure}
  \hspace*{\algorithmicindent} \textbf{Input:} graph $G$\\
  \hspace*{\algorithmicindent} \textbf{Output:} rank $R_{\text{SCC}}$, $R_{\text{IN}}$, $R_{\text{OUT}}$, $R_{\text{TUBES}}$, $R_{\text{INTENDRILS}}$, $R_{\text{OUTTENDRILS}}$, $R_{\text{OTHERS}}$
  \begin{algorithmic}[1]
  \State Generate $1000$ random graphs $G^{*}_{i}$, $i = 1,2,..., 1000$ based on graph $G$, using Newman et al.'s directed configuration model \cite{dcm}.
  \State For each random graph $G^{*}_{i}$, acquire its bow-tie structure $\text{SCC}^{*}_{i}$, $\text{IN}^{*}_{i}$, $\text{OUT}^{*}_{i}$, $\text{TUBES}^{*}_{i}$, $\text{INTENDRILS}^{*}_{i}$, $\text{OUTTENDRILS}^{*}_{i}$, $\text{OTHERS}^{*}_{i}$. Do the same for graph $G$ to obtain SCC, IN, OUT, TUBES, INTENDRILS, OUTTENDRILS, OTHERS.
  \State Calculate the rank of SCC $R_{\text{SCC}} = \frac{\sum_{i = 1}^{1000}\mathbbm{1}_{\lvert \text{SCC}^{*}_{i}\rvert < \lvert \text{SCC} \rvert}}{1000}$\footnotemark. Repeat this step to calculate the rank of other bow-tie components $R_{\text{IN}}$, $R_{\text{OUT}}$, $R_{\text{TUBES}}$, $R_{\text{INTENDRILS}}$, $R_{\text{OUTTENDRILS}}$, $R_{\text{OTHERS}}$.  \end{algorithmic}
\end{algorithm}
\footnotetext{For any set $A$, we define $\rvert A \lvert$ as the cardinality of $A$, that is, the number of elements in $A$.}

\begin{figure}[bpht]
\centering
\adjincludegraphics[width=\textwidth,trim={{.04\width} {.08\height} {.04\width} {.07\height}},clip]{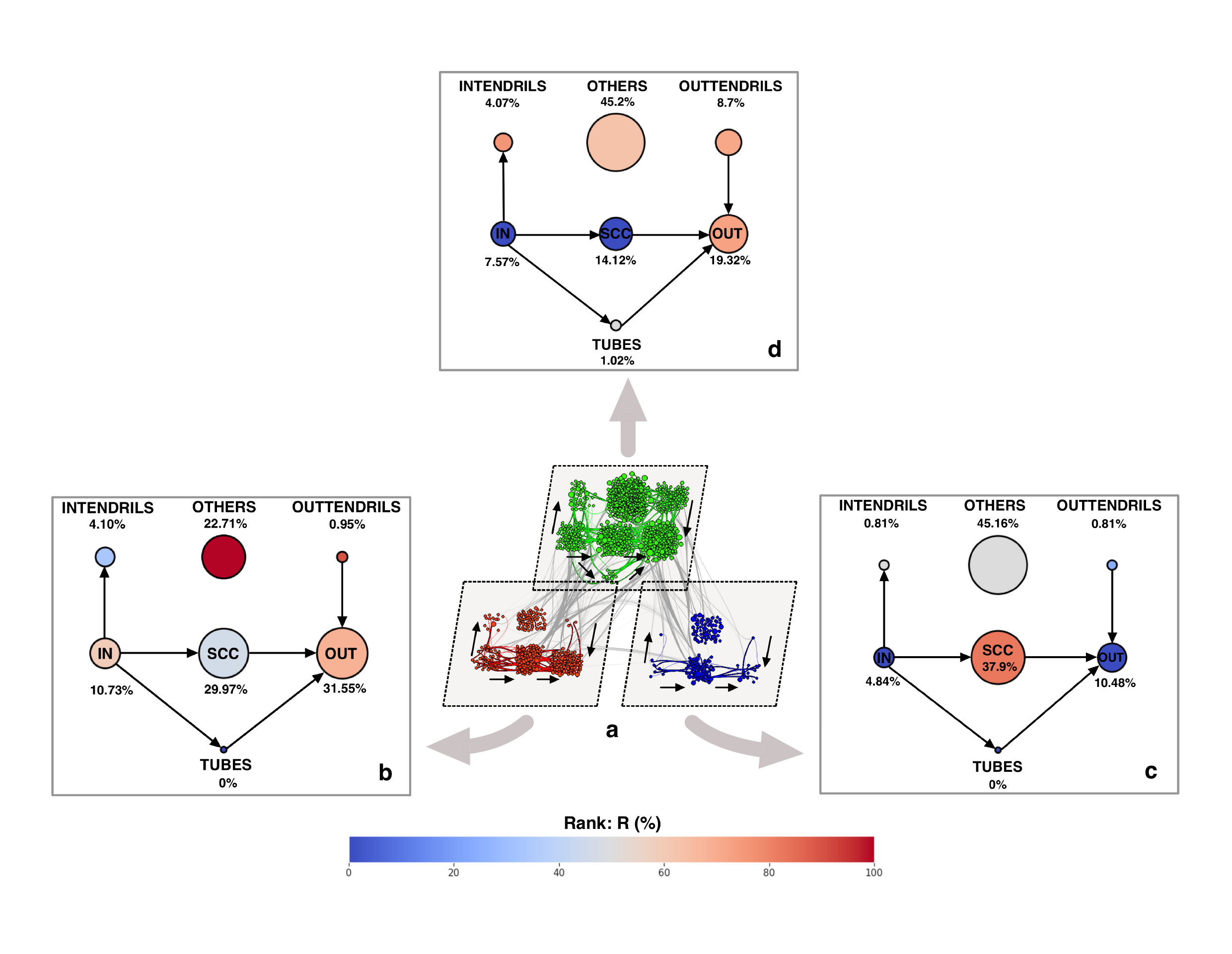}
\caption{\textbf{Statistical significance of the within-group bow-tie decomposition in the February 2019 online recommendation-based network. }
\textbf{a, implementation scheme. } A replica of the diagram in Figure 3 in our main paper, the recommendation-based network in February is partitioned into subgraphs according to vaccination groups (red for anti, blue for pro, green for neutral).
\textbf{b, c, d, quantitative results for each bow-tie structure.} 
Bow-tie structure of each subgraph is shown using an organised layout of node and arrow, maintaining consistency with Figure 1b in our main paper\protect\footnotemark. 
Node size and numeric values indicate the proportion of pages held by each bow-tie component in each vaccine group. 
Node colour indicates the rank $R$ of each bow-tie component among 1000 graphs generated by edge reorganisation while maintaining fixed in-degree and out-degree sequences in the observed graph.
This $R$ represents the percentage of random graphs in which the corresponding bow-tie component is smaller than the observed graph. A high or low $R$ value, represented by darker shared of red or blue, indicates statistical significance in the bow-tie component size compared to the observed graph.
By observation, the large components detected in our dataset (i.e., SCC in the pro-vaccination group, OUT in the anti-vaccination group, and OTHERS in the neutral group) are displayed in red. 
This indicates that, in many randomly generated graphs, these components are smaller than those in our dataset, suggesting that they are not solely the result of degree distributions.
}
\label{fig:sig-within}
\end{figure}

\begin{figure}[bpht]
\centering
\adjincludegraphics[width=\textwidth,trim={{.01\width} {.01\height} {.01\width} {.01\height}},clip]{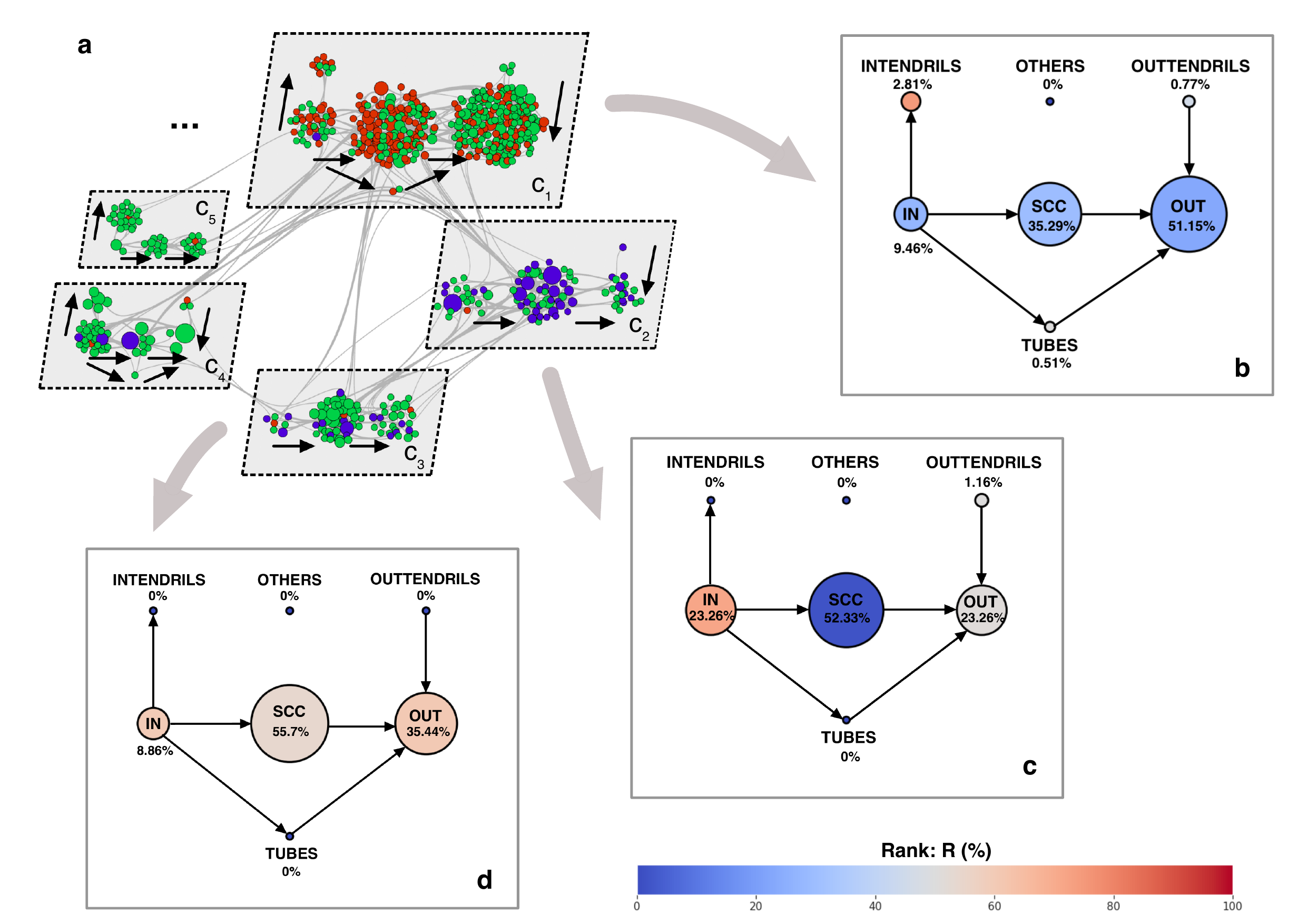}
\caption{\textbf{Statistical significance of the across-group bow-tie decomposition in the February 2019 online recommendation-based network.} 
All sub-figures are presented in the same format as Figure \ref{fig:sig-within}, except that the detected communities are the targeted subgraphs instead of the vaccination groups. 
Our results show that the observed large components (e.g. OUT in $C_1$ and SCC in $C_2$) are generally smaller than those in the corresponding random graphs, as indicated by the presence of blue colours in such components.
This suggests the statistical significance of the across-group bow-tie structure, where pages tend not to form large bow-tie components (which is
interestingly contrary to the case in within-group bow-tie structure).
Moreover, the more neutral pages a community contains, the lower statistical significance the bow-tie structure of this community manifests (e.g., community $C_3$). 
These observations could be attributed to neutral pages having fewer vaccination-related communication purposes and exhibiting more varied behaviour.
}
\label{fig:sig-across}
\end{figure}

\paragraph{Results.}
As illustrated in Figure \ref{fig:sig-within} and \ref{fig:sig-across}, both within-group and across-group bow-tie decomposition demonstrate strong statistical significance, though interestingly with some differences. 
As to the within-group bow-tie decomposition, the rank of each bow-tie component generally confirms the spontaneity of forming large components for all three vaccination groups. For instance, the red colour of the SCC component in the pro-vaccination group implies that a large number of random graphs generate SCC that are smaller than the observed one, thereby indicating that the observed large SCC does not solely result from the degree distribution. The same applies to the OUT component in the anti-vaccination group and the OTHERS component in the neutral group. 
Conversely in the across-group bow-tie decomposition, pages generally don't form 
such large bow-tie components. That is, the observed large components are generally smaller than those in the corresponding random graphs, as indicated by the presence of blue colours in such components. This could be attributed to the participation of neutral pages in identified communities, as they have fewer vaccination-related communication purposes and may exhibit more varied 
behaviour. 

\footnotetext{Note that Figure \ref{fig:sig-within}b, c, d  also resemble figures in Mattei et al. \cite{Mattei-2022}, enabling easy comparison.}

\subsubsection*{B.2. Follow-up Dataset}

\begin{figure}[bpht]
\centering
\adjincludegraphics[width=\textwidth,trim={{.01\width} {.01\height} {.01\width} {.04\height}},clip]{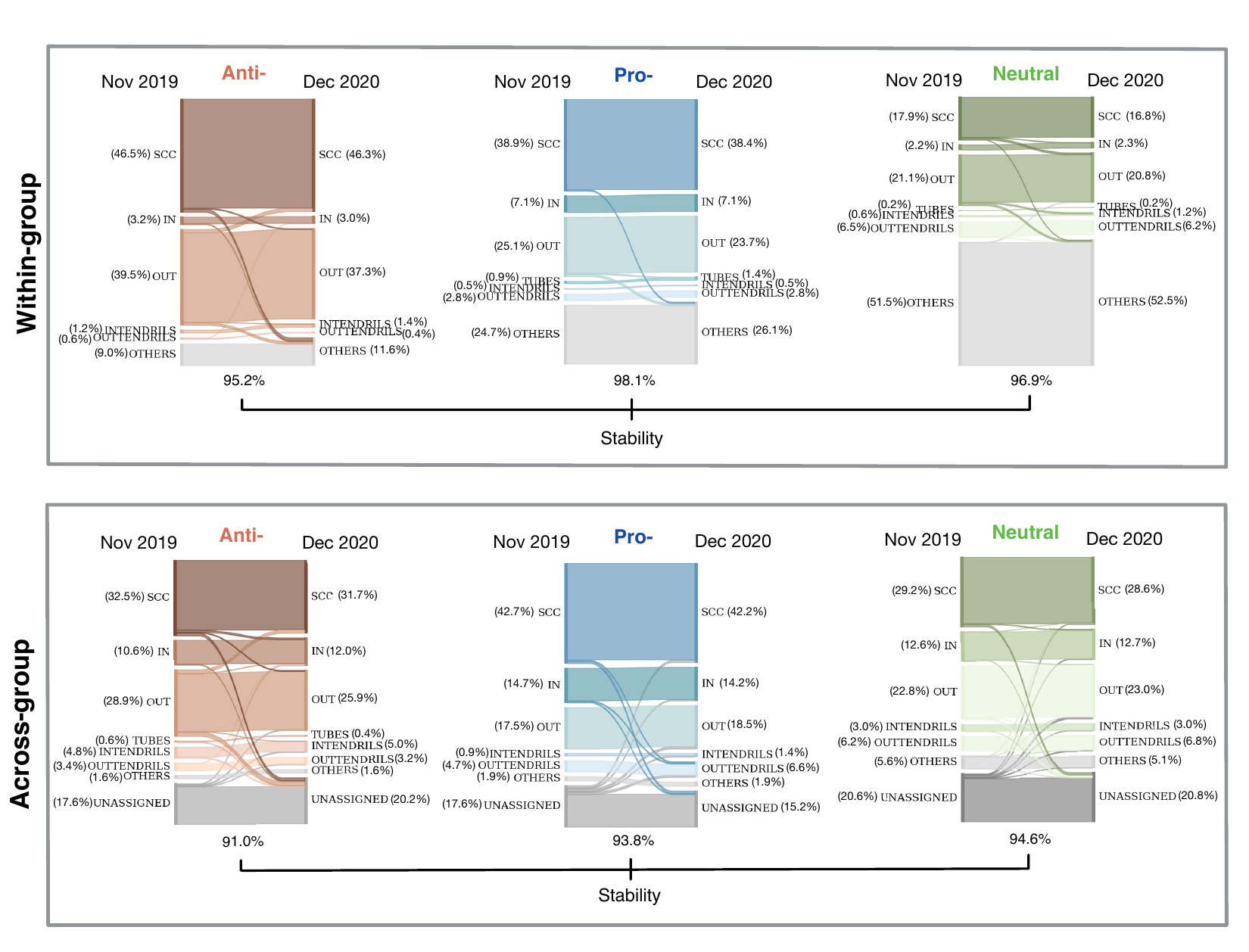}
\caption{\textbf{Within-group and across-group bow-tie structures in the November 2019 and December 2020 online recommendation-based networks.} These Sankey diagrams are structured in the same format as Figure 3 in our main paper for the purpose of comparison between the main dataset and the follow-up dataset. 
The obtained results generally align with our main dataset findings: bow-tie structures of pro-vaccination pages have a relatively large SCC component, both within-group and across-group, while anti-vaccination pages have a large OUT component. In contrast, the neutral pages have a large OTHERS in within-group bow-tie decomposition, but demonstrate a dissimilar pattern in across-group bow-tie decomposition with comparatively large main bow-tie components (i.e., SCC, OUT, and IN). 
Nevertheless, the temporal stability of bow-tie structures for pages in all three vaccination groups appears significantly higher than in the main dataset, and neutral pages no longer exhibit the lowest stability.}
\label{fig:stability-update}
\end{figure}

The follow-up dataset from Illari et al. \cite{Illari-2022} contains another two snapshots: one from November 2019 and the other from December 2020, during the early stage of the COVID-19 pandemic. Both snapshots involve identical 1356 nodes, with 7387 edges in November 2019 and 7154 edges in December 2020.
Its data framework remains consistent with the main dataset in our paper, with the exception that in the December 2020 snapshot, the edges only represent recommendations occurring between November 2019 and December 2020, rather than prior to December 2020. In this sense, edges in the November 2019 snapshot do not necessarily appear in the December 2020 snapshot. 

This follow-up dataset has some limitations in the context of our bow-tie analysis. 
Firstly, the $1356$ nodes in this dataset do not entirely match the $1326$ nodes in our main dataset, and these $1356$ nodes lack unique IDs for establishing correspondence between nodes in the two datasets.
Secondly, only page fan counts for November 2019 are available; there is no such data for December 2020. This absence of dynamic fan size data hinders our analysis of opinion dynamics and affects our Infomap community detection results, which depend on edge weights derived from the product of both ends' fan size.

Despite these limitations, recursive bow-tie decomposition remains viable in the follow-up dataset. Note that, for Infomap implementation purposes, we have organised the pages in the December 2020 snapshot with the same fan counts as the November 2019 snapshot.
As shown in Figure \ref{fig:stability-update}, our findings indicate that all three vaccination groups in the follow-up dataset exhibit similar bow-tie structures to the main dataset. 
Nevertheless, the temporal stability of these bow-tie structures appears significantly higher than in the main dataset, and neutral pages no longer exhibit the lowest stability. This disparity may stem from the fixed fan sizes at both timestamps and the difference in edge construction methods between the two datasets.


\subsection*{C. Supervised Machine Learning: Feature Investigation}

\begin{figure}[bpht]
    \centering
    \adjincludegraphics[width=0.6\textwidth,trim={{.01\width} {.01\height} {.01\width} {.01\height}},clip]{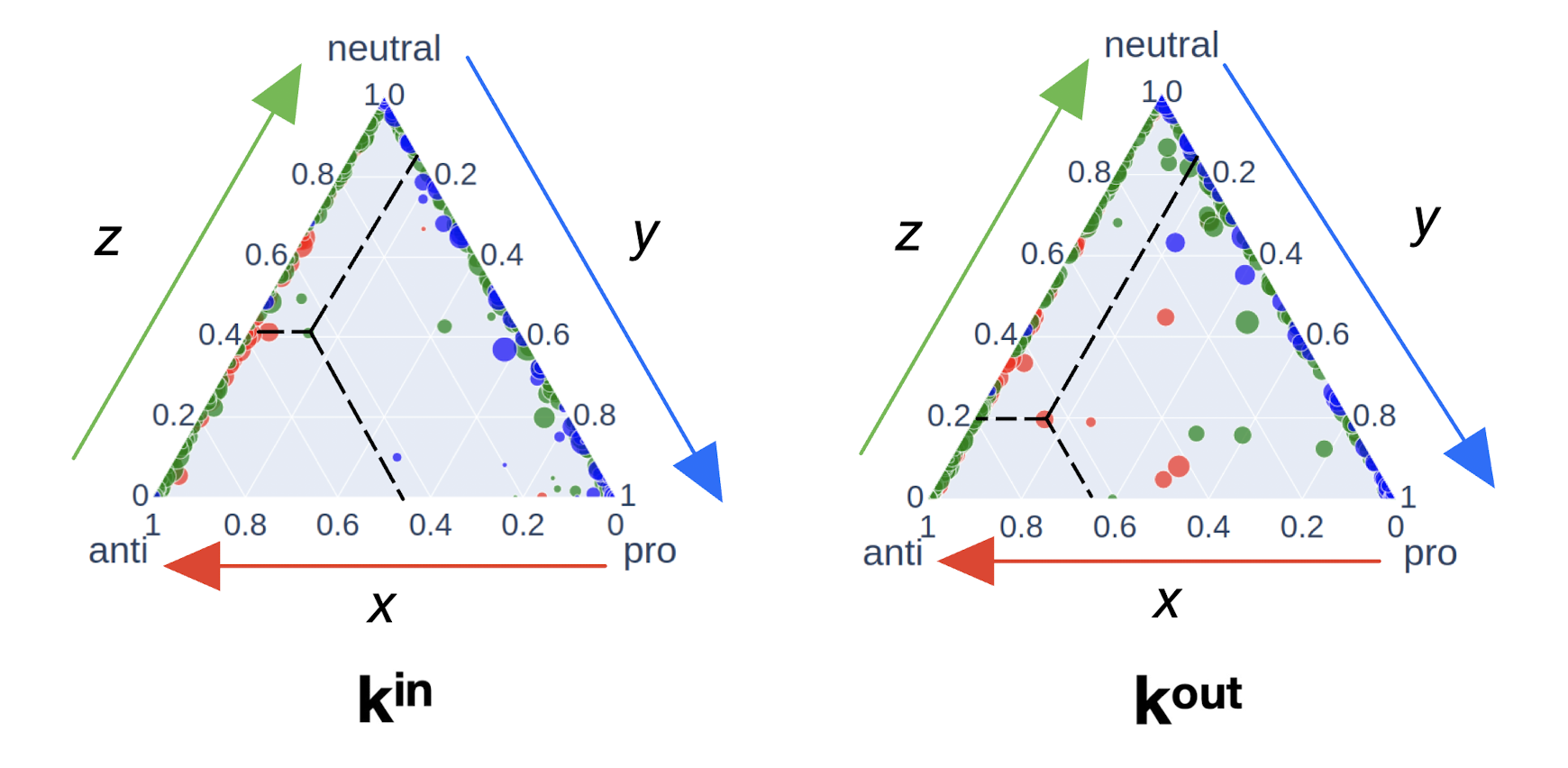}
    \caption{\textbf{Weighted in-degree (\bm{$k^{in}$}) and out-degree (\bm{$k^{out}$}) composition in relation to polarity.}
    Each page $i$ is represented by a node in both triplots, coloured to show page polarity (red for anti, blue for pro, green for neutral), with size proportional to its fan count in the February 2019 snapshot.
    The coordinates $(x_i^{in}, y_i^{in}, z_i^{in}):= (\frac{\sum_{j} \mathbbm{1}_{p_i = r} A_{ji}}{k_i^{in}}, \frac{\sum_{j} \mathbbm{1}_{p_i = b} A_{ji}}{k_i^{in}} , \frac{\sum_{j} \mathbbm{1}_{p_i = g} A_{ji}}{k_i^{in}})$ and $(x_i^{out}, y_i^{out}, z_i^{out}) := (\frac{\sum_{j} \mathbbm{1}_{p_i = r} A_{ij}}{k_i^{out}}, \frac{\sum_{j} \mathbbm{1}_{p_i = b} A_{ij}}{k_i^{out}} , \frac{\sum_{j} \mathbbm{1}_{p_i = g} A_{ij}}{k_i^{out}})$ in the two triplots indicate the composition of its weighted recommendations from and to other pages, based on page polarity (notations follow Table 1 in our main paper). Nodes located on vertices suggest pages' absolute inclination towards a polarity.
    Most nodes are found along the y and z axes in both triplots, with a slightly higher concentration of nodes in the central area evident in the $k^{out}$ triplot. 
    Specifically, green nodes appear on both the y and z axes, showing neutral pages align one-sidedly when interacting (i.e., whether through recommendations or being recommended) with non-neutral ones. Red nodes are mainly on the z-axis, while blue nodes are on the y-axis, illustrating that anti-vaccination pages primarily interact with like-minded or neutral pages. 
    Similarly, this trend persists for pro-vaccination pages.}
    \label{fig:degree}
\end{figure}

When extracting features of anti- and pro-vaccination pages from the February 2019 recommendation-based network snapshot to predict their fan count variations compared with October 2019, we solely specify the features indicating the proportion of within-vaccination-group page interactions (i.e., $\textit{k-PS}_i^{in}$ and $\textit{k-PS}_i^{out}$), and we do not delineate features indicating the proportion of inter-vaccination-group page interactions with specific vaccination groups.
That is because we find that anti- (pro-) pages predominantly engage with similar-minded or neutral pages, rarely with pro- (anti-) ones. Moreover, neutral pages seldom reciprocate recommendations with both anti- and pro-vaccination pages. Instead, if neutral pages interact (i.e., recommend or are recommended) with non-neutral counterparts, a clear one-sided leaning emerges. These trends are visually shown in Figure \ref{fig:degree}. 

\subsection*{D. Agent-based SIR Model: Parameter Choices}

\begin{figure}[bpht]
    \centering
    \adjincludegraphics[width=0.9\textwidth,trim={{.01\width} {.33\height} {.01\width} {.22\height}},clip]{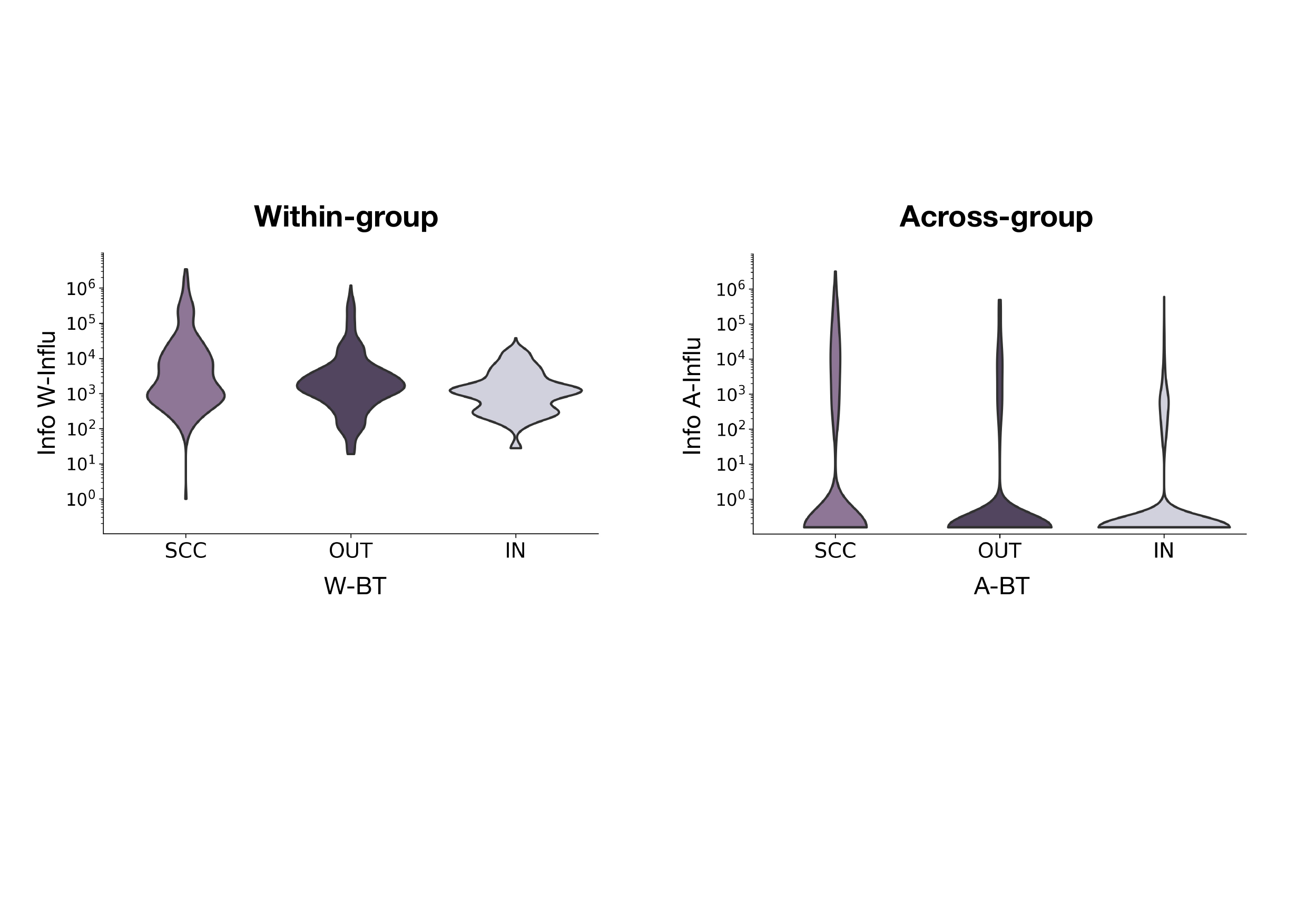}
    \caption{\textbf{Distinctions among the influence of information pieces initialised from different bow-tie components.} 
    The figure layout is consistent with Figure 4 from our main paper, with $\beta = 0.3$ and $\gamma = 0.2$ here. 
    }
    \label{fig:sisir1}
\end{figure}

\begin{figure}[ht!]
    \centering
    \adjincludegraphics[width=\textwidth,trim={{.03\width} {.14\height} {.03\width} {0.03\height}},clip]{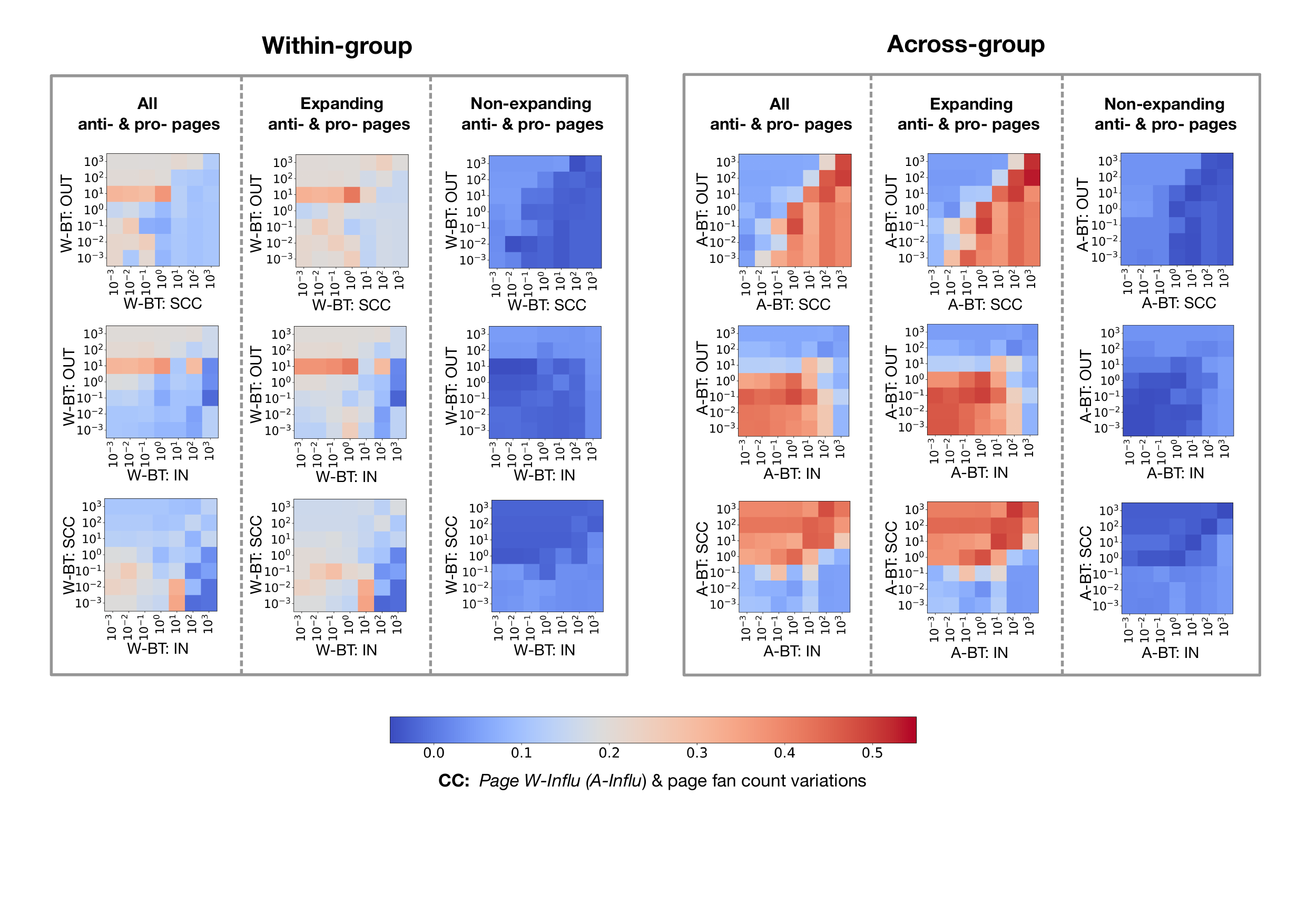}
    \caption{\textbf{Distinctions among the influence of pages in different bow-tie components when varying their probability of generating information pieces.} 
    The figure layout is consistent with Figure 5 from our main paper, with $\beta = 0.3$ and $\gamma = 0.2$ here. 
    }
    \label{fig:sisir2}
\end{figure}

The simulation of information cascades in our agent-based SIR model is required to choose the transmission rate $\beta$ and recovery rate $\gamma$. 
The influence of each information piece $n$ (i.e., $\textit{Info W-Influ}_n$ and $\textit{Info A-Influ}_n$) depends on the effective infection ratio $\frac{\beta}{\gamma}$ \cite{Cinelli-2021}. 
In our paper, we do not distinguish between individual information pieces, but rather focus on the volume of information generated by each page. Therefore, we use the same values of $\beta$ and $\gamma$ for all generated information pieces.
We fix $\beta = 0.5$ and $\gamma =0.3$ for the results presented in our main paper. Here we show that our results are robust for different parameter choices $\beta = 0.3$ and $\gamma = 0.2$ (see Figure \ref{fig:sisir1} and \ref{fig:sisir2}). 

We acknowledge that not all choices of $\beta$ and $\gamma$ can produce similar results, and we make the example parameter choices above considering the rationality of our simulation. 
For example, the effective infection ratio $\frac{\beta}{\gamma}$ needs to strike an appropriate balance, avoiding excessive values that either spread information too widely, or restrict it primarily to neighbourhood pages, mostly sharing the source page's vaccination stance.

\end{document}